\RequirePackage{fix-cm}
\documentclass[twocolumn]{svjour3}          
\smartqed  

\usepackage{color}
\usepackage{booktabs}
\usepackage{multirow}
\usepackage{array}
\usepackage{lipsum}
\usepackage{graphicx}
\usepackage{caption}
\usepackage{amsmath}
\usepackage[colorlinks, linkcolor=red, anchorcolor=blue, citecolor=green]{hyperref}
\usepackage{float}
\usepackage{subfigure}
\usepackage{rotating}
\usepackage{makecell}
\usepackage[abs]{overpic}
\usepackage[noend]{algpseudocode}
\usepackage{algorithmicx,algorithm}
\usepackage{amssymb}

\newenvironment{shrinkeq}[1]
{ \bgroup
	\addtolength\abovedisplayshortskip{#1}
	\addtolength\abovedisplayskip{#1}
	\addtolength\belowdisplayshortskip{#1}
	\addtolength\belowdisplayskip{#1}}
{\egroup\ignorespacesafterend}

\def\eg{\emph{e.g.}}

\definecolor{lvgreen}{rgb}{0.2,0.6,0.15}
\definecolor{lvblue}{rgb}{0,0.3,0.9}
\begin{document}

\title{Attention Guided Low-light Image Enhancement with a Large Scale Low-light Simulation Dataset}
\author{Feifan Lv         \and
	Yu Li         \and
	Feng Lu 
}


\institute{F. Lv \at
	Beihang University, Beijing, China.
	\and
	Y. Li \at
	Applied Research Center (ARC), Tencent PCG, Shenzhen, China.
	\and
	F. Lu (Corresponding author)\at
	Beihang University, Beijing, China \\
	Peng Cheng Laboratory, Shenzhen, China \\
	\email{lufeng@buaa.edu.cn}
}

\date{Received: date / Accepted: date}

\maketitle

\begin{abstract}
	Low-light image enhancement is challenging in that it needs to consider not only brightness recovery but also complex issues like color distortion and noise, which usually hide in the dark. Simply adjusting the brightness of a low-light image will inevitably amplify those artifacts. To address this difficult problem, this paper proposes a novel end-to-end attention-guided method based on multi-branch convolutional neural network. To this end, we first construct a synthetic dataset with carefully designed low-light simulation strategies. The dataset is much larger and more diverse than existing ones. With the new dataset for training, our method learns two attention maps to guide the brightness enhancement and denoising tasks respectively. The first attention map distinguishes underexposed regions from well lit regions, and the second attention map distinguishes noises from real textures. With their guidance, the proposed multi-branch decomposition-and-fusion enhancement network works in an input adaptive way. Moreover, a reinforcement-net further enhances color and contrast of the output image. Extensive experiments on multiple datasets demonstrate that our method can produce high fidelity enhancement results for low-light images and outperforms the current state-of-the-art methods by a large margin both quantitatively and visually.
\end{abstract}

\section{Introduction}
\label{intro}
Images captured in insufficiently illuminated environment usually contain undesired degradations, such as poor visibility, low contrast, unexpected noise, etc. Resolving these degradations and converting low-quality low-light images to normally exposed high-quality images require well developed low-light enhancement techniques. Such a technique has a wide range of applications. For example, it can be used in consumer photography to help the users capture appealing images in the low-light environment. It is also useful for a variety of intelligent systems, \eg, automated driving and video surveillance, to capture high-quality inputs under low-light conditions.

\begin{figure*}[htbp]
	\begin{center}
		\begin{overpic}[width=1\textwidth]{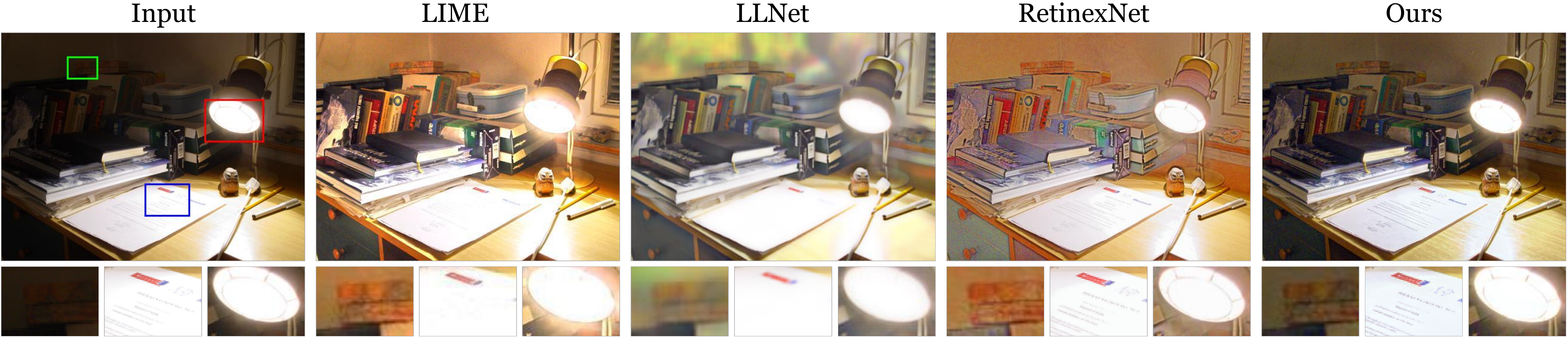}				
			\put(158,100){\bf \color{black}\footnotesize \cite{guo2017lime}} 
			\put(258,100){\bf \color{black}\footnotesize \cite{lore2017llnet}} 
			\put(368,100){\bf \color{black}\footnotesize \cite{Chen2018Retinex}} 
		\end{overpic}
	\end{center}
	\caption{Low-light enhancement example. Comparing with existing methods, our method can generate results with satisfactory visibility, natural color, and higher contrast.}
	\label{fig_first}
\end{figure*}

Low-light image enhancement is still a challenging task, since it needs to manipulate color, contrast, brightness and noise simultaneously given the low quality input only. Although numbers of methods have been proposed for this task in recent years, there is still large room for improvement. Figure~\ref{fig_first} shows some limitations of existing methods, which follow typical assumptions of histogram equalization (HE) and Retinex theory~\cite{land1977retinex}. HE-based methods aim to increase the contrast by simply stretching the dynamic range of images, while Retinex-based methods recover the contrast by using the estimated illumination map. Mostly, they focus on restoring brightness and contrast and ignore the influences of noise. However, in reality, the noise is inevitable and non-negligible in the low-light images, especially after increasing brightness and contrast.

To suppress the low-light image noise, some methods directly include a denoising process as a separate component in their enhancement pipeline. However, it is dilemma to make a simple cascade of the denoising and enhancement procedures. In particular, applying denoising before enhancement will result in blurring, while applying enhancement before denoising will cause noise amplification. Therefore, in this paper, we propose to model and solve the denoising and low-light enhancement problems simultaneously.

Specifically, this paper proposes an attention-guided enhancement solution that achieves denoising and enhancing simultaneously and effectively. We find that the severity of low brightness/contrast and high image noise show certain spatial distributions related to the underexposed areas. Therefore, the key is to handle the problem in a region-aware adaptive manner. To this end, we propose the under-exposed (ue) attention map to evaluate the degree of underexposure. It guides the method to pay more attention to the underexposed areas in low light enhancement. In addition, based on the ue-attention map, we derive the noise map to guide the denoising according to the joint distribution of exposure and noise intensity. Subsequently, we design a multi-branch CNN to simultaneously achieve low-light enhancement and denoising under the guidance of both maps. In the final step, we add a fully-convolutional network for improving the image contrast, exposure and color as the second enhancement.

The remaining difficulty lies in the lack of large-scale paired low-light image dataset, making it challenging to train an effective network. To address this issue, we propose a low-light image simulation pipeline to synthesize realistic low-light images with well exposed ground truth images. Image contrast and color are also improved to provide good references for our image re-enhancement step. Following the above ideas, we propose a large-scale low-light image dataset as an efficient benchmark for low-light enhancement researches.

Overall, our contributions are in three folds:
1) We propose a full pipeline for low-light image simulation with high fidelity, based on which we build a new large-scale paired low-light image dataset to support low-light enhancement researches.
2) We propose an attention-guided enhancement method and the corresponding multi-branch network architecture. Guided by the ue-attention map and noise map, the proposed method achieves low-light enhancement and denoising simultaneously and effectively.
3) Comprehensive experiments have been conducted and the experiment results demonstrate that our method outperforms state-of-the-art methods by a large margin.


\section{Related Work}
Image enhancement and denoising have been studied for a long time. In this section, we will briefly overview the most related methods.

{\bf Traditional enhancement methods.} Traditional methods can be mainly divided into two categories. The first category is built upon the histogram equalization (HE) technique. The differences of different HE-based methods are using different additional priors and constraints. In particular, BPDHE~\cite{ibrahim2007bpdhe} tries to preserve image brightness dynamically; Arici~\textit{et al}.~\cite{arici2009wahe} propose to analyze and penalize the unnatural visual effects for better visual quality; DHECI~\cite{nakai2013dheci} introduces and uses the differential gray-level histogram; CVC~\cite{celik2011contextual} uses the interpixel contextual information; LDR~\cite{lee2013contrast} focuses on the layered difference representation of 2D histogram to try to enlarge the gray-level differences between adjacent pixels. These methods expand the dynamic range and focus on improving the contrast of the entire image instead of considering the illumination. They may cause the problem of over- and under-enhancement.

The other category is based on the Retinex theory~\cite{land1977retinex}, which assumes that an image is composed of reflection and illumination. Typical methods, \textit{e.g.}, MSR~\cite{jobson1997multiscale} and SSR~\cite{jobson1997properties}, try to recover and use the illumination map for low-light image enhancement. Recently, AMSR~\cite{lee2013amsr} proposes a weighting strategy based on SSR. NPE~\cite{wang2013naturalness} balances the enhancement level and image naturalness to avoid over-enhancement. MF~\cite{fu2016mf} processes the illumination map in a multi-scale fashion to improve the local contrast and maintain naturalness. SRIE~\cite{fu2016srie} develops a weighted vibrational model for illumination map estimation. LIME~\cite{guo2017lime} develops a structure-aware smoothing model to estimate the illumination map. BIMEF~\cite{ying2017bio} proposes a dual-exposure fusion algorithm and Ying~\textit{et al}.~\cite{ying2017newiccv} use the camera response model for further enhancement. Mading~\textit{et al}.~\cite{li2018structure} propose a robust Retinex model by considering the noise map for enhancing low-light images accompanied by intensive noise. However, the key to these Retinex-based methods is the estimation of the illumination map, which is hand-crafted and relied on careful parameters tuning. Besides, most of these Retinex-based methods do not consider noise removal and often amplify the noise.

{\bf Learning-based enhancement methods.}
Recent-ly, deep learning has achieved great success in the field of low-level image processing~\cite{sharma2018classification} and
nighttime scenes modeling~\cite{radenovic2016dusk} and understanding~\cite{daytime:2:nighttime,GCMA:UAE:NighttimeSegmentation19}.
Powerful tools such as end-to-end networks and GANs~\cite{goodfellow2014generative} have been used in image enhancement.
LLNet~\cite{lore2017llnet} uses the multilayer perception auto-encoder for low-light image enhancement and denoising.
HDRNet~\cite{gharbi2017deep} learns to make local, global, and content-dependent decisions to approximate the desired image transformation. LLCNN~\cite{tao2017llcnn} and~\cite{tao2017low} rely on some traditional methods and are not end-to-end solutions to handle brightness/contrast enhancement and denoising simultaneously. MSRNet~\cite{shen2017msr} learns an end-to-end mapping between dark/bright images by using different Gaussian convolution kernels.
MBLLEN~\cite{lvmbllen} uses a novel multi-branch low-light enhancement network architecture to learn the mapping from low-light images to normal light ones.
Retinex-Net~\cite{Chen2018Retinex} combines the Retinex theory with CNN to estimate the illumination map and enhance the low-light images by adjusting the illumination map. Similarly, KinD~\cite{zhang2019kindling} designs a similar network by adding a Restoration-Net for noise removal. Ren~\textit{et al}.~\cite{ren2019low} propose a novel hybrid network contains a content stream and a salient edge stream for low-light image enhancement. DeepUPE~\cite{wang2019underexposed} proposes a network for enhancing underexposed images by estimating an image-to-illumination mapping. However, it does not consider the low-light noise.

Besides, DPED~\cite{ignatov2017dslr,de2018fast} proposes an end-to-end approach using a composite perceptual error function for translating low-quality mobile phone photos into DSLR-quality photos. PPCN~\cite{Hui-PPCN-2018} designs a compact network and combines teacher-student information transfer to reduce computational cost. WESPE~\cite{ignatov2018wespe} proposes a weakly-supervised method to overcome the restrictions on requiring paired images. Also, Chen~\textit{et al}.~\cite{chen2018deep} propose an unpaired learning method for
image enhancement by improving two-way GANs.
As for extremely low-light scenes, SID~\cite{seedark2018cvpr} develops a CNN-based pipeline to directly process raw sensor images.
Lv~\textit{et al}.~\cite{lv202024hour} propose an enhancement solution by separating the visible and near-infrared signal from a single image and fusing them for high-quality images.
Most of these learning-based methods do not explicitly contain the denoising process, and some even rely on traditional denoising methods. However, our approach considers the effects of noise and uses two attention maps to guide the enhancing and denoising process. So, our method is complementary to existing learning-based methods.

{\bf Image denoising methods.} Existing works for image denoising are massive. For Gaussian denoising, BM3D~\cite{dabov2006image} and DnCNN~\cite{zhang2017beyond} are representatives of the filter-based and deep-learning-based methods. For Poisson denoising, NLPCA~\cite{salmon2014poisson} combines elements of dictionary learning with sparse patch-based representations of images and employs an adaptation of Principal Component Analysis. Azzari~\textit{et al}.~\cite{azzari2016variance} propose an iterative algorithm combined with variance-stablizing transformation (VST) and BM3D filter~\cite{dabov2006image}. DenoiseNet~\cite{remez2017deep} uses a deep convolutional network to calculate the negative noise components, which adds directly to the original noisy image to remove Poisson noise. For Gaussian-Poisson mixed denoising, CBDNet~\cite{Guo2019Cbdnet} presents a convolutional blind denoising network by incorporating asymmetric learning. It is applicable to real noise images by training on both synthetic and real images. For real-world image denoising, TWSC~\cite{TWSC_ECCV2018} develops a trilateral weighted sparse coding scheme. Chen~\textit{et al}.~\cite{chen2018image} propose a two-step framework which contains noise distribution estimation using GANs and denoising using CNNs. Directly combining these methods with enhancement methods will result in blurring. To avoid this, our solution performs enhancing and denoising simultaneously.

\begin{figure*}[t]
	\begin{center}
		\includegraphics[width=1\textwidth]{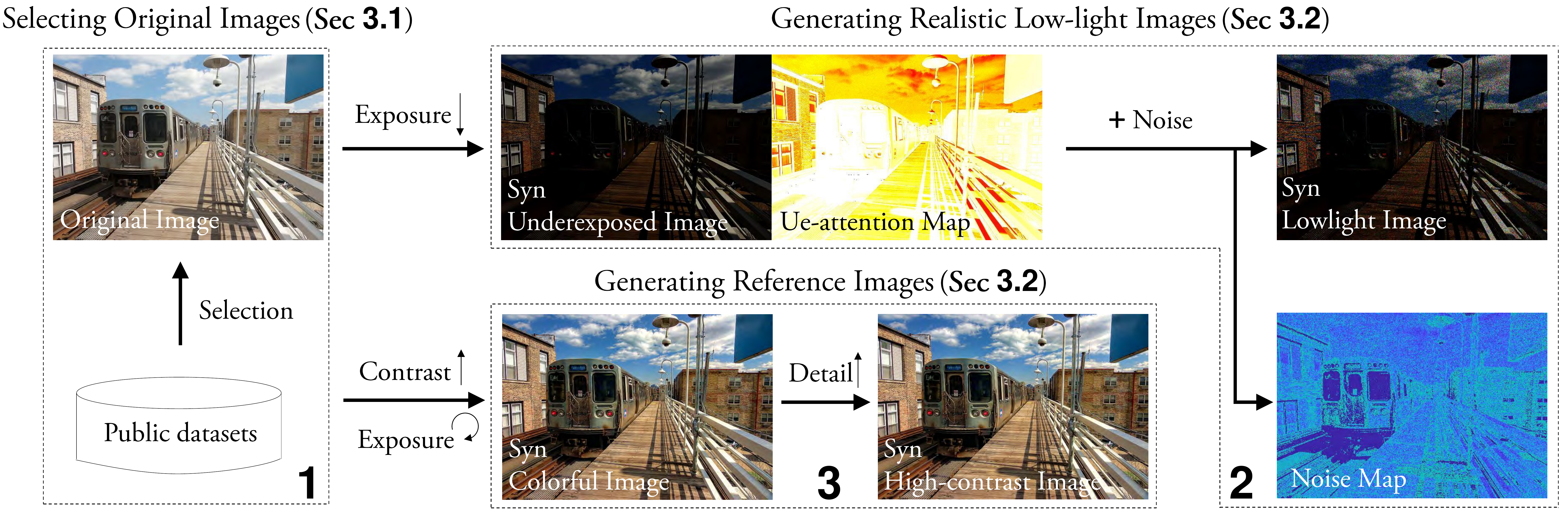}
	\end{center}
	\caption{Pipeline of constructing the proposed low-light simulation dataset. Our method optimally selects normal exposed images from public datasets, performs low-light simulation, and adds noise to synthesize realistic low-light images. Meanwhile, the original normal exposed images are enhanced by exposure correction and contrast/details amplification, so as to generate high-quality reference images. Details can be found in Section~\ref{sec_dataset}.}
	\label{fig_process}
\end{figure*}

{\bf Low-light Image Enhancement Datasets.}
Several previous datasets are constructed by manually capturing paired low-light and normal-light images. Multiple shootings with different camera configurations or retouching captured images are the two main solutions.
The LOL~\cite{Chen2018Retinex} and SID~\cite{seedark2018cvpr} datasets are constructed using the former solution.
Images in the LOL dataset are captured in the daytime by controlling the exposure and ISO. Meanwhile, the underexposed images are generated by linear degradation approximately, which may differ from real cases. This will result in performance variation in low-light image enhancement (see the result of RetinexNet in Figure~\ref{fig_first}). The SID dataset is composed of raw sensor data under extremely low-light scenes, which is different from those used in general low-light image enhancement researches.
As for the latter solution, the DeepUPE~\cite{wang2019underexposed} dataset collects 3,000 underexposed images, each with an expert-retouched reference. However, the under-exposed levels of the images are relatively low, which may not cover the heavily low-light scenes. Besides, the SICE~\cite{Cai2018deep} dataset collects multi-exposed image sequences and uses the Exposure Fusion methods to construct the reference image under the supervision of human. However, imperfect alignment of image sequences will result in blur and ghosting.
Although these datasets have made great contributions to the field of low-light image enhancement, they still show limitations.
On one hand, their data amounts are relatively small with respect to the number of images. Since the variation of scenes and light conditions are limited, the trained models may not be generalized well in many cases.
On the other hand, due to the lack of annotations, these datasets are difficult to be used for other relevant vision tasks, such as detection and segmentation in the dark.

\section{Large Scale Low-Light Simulation Dataset}
\label{sec_dataset}
In this paper, we propose an effective low-light simulation method to synthesize low-light images from normal-light images. The purpose is to offer a large diversity in scenes and light conditions which is required by our method and other further researches. Many previous works~\cite{FoggySynscapes,semantic:foggy:scene} have proven that the synthetic data is an effective alternative to real data in different vision tasks. Using synthetic data allows easy model adaptation for target conditions without requiring additional manual annotations~\cite{SynRealDataFog19,SynRealDataFogECCV18}. Similarly, we believe that generating synthetic low-light image datasets from public datasets~\cite{bileschi2006streetscenes,everingham2010pascal,grubinger2006iapr,lin2014microsoft} with rich annotations also has the potential to achieve model adaptation in low-light conditions. The proposed dataset construction pipeline is shown in Figure~\ref{fig_process}.

\begin{figure}[t]
	\begin{center}
		\begin{overpic}[width=0.48\textwidth]{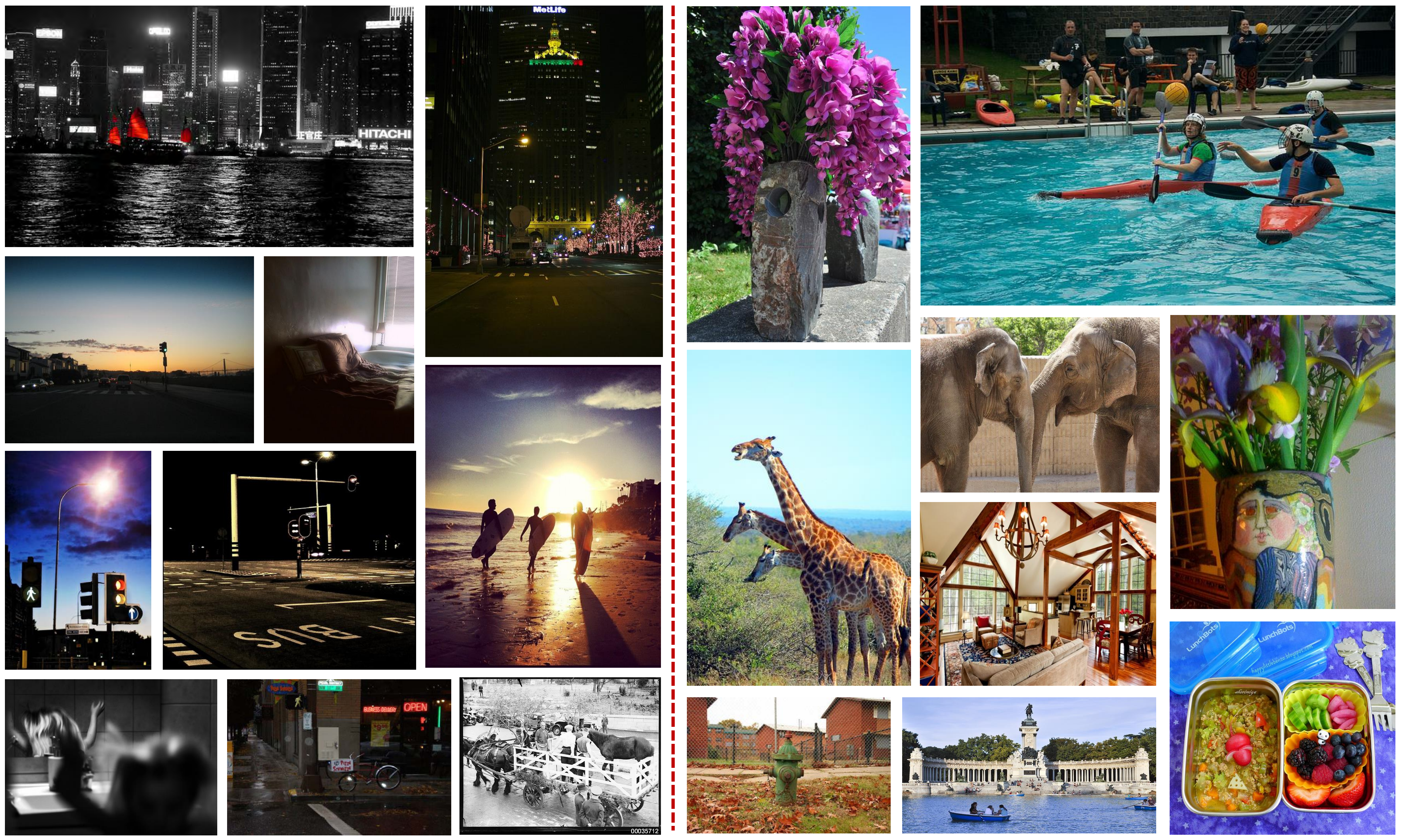}
		\end{overpic}
	\end{center}
	\caption{Samples of large-scale public datasets: (left) low-quality examples, (right) high-quality examples.}
	\label{fig_second}
\end{figure}

\subsection{Candidate Image Selection}
Our proposed low-light simulation requires high-quality normally exposed images as input, and these images also serve as the reference for low-light enhancement. Therefore, we need to distinguish such high-quality images from low-quality ones given large-scale public image datasets, as shown in Figure~\ref{fig_second}. To this end, we propose a candidate image selection method which takes the proper exposure, rich color, blur-free and rich details into account. The selection method contains three steps: darkness estimation, blur estimation and color estimation.

{\bf Darkness estimation.} To select images with sufficient exposure, we first apply over-segmentation~\cite{achanta2012slic} and restore the segmentation results. Subsequently, we calculate the mean/variance of the \textit{V} component in \textit{HSV} color space based on the segmentation results. If the calculated mean/variance is larger than thresholds, we set this segmentation block to be sufficiently exposed. Finally, images with more than $85\%$ sufficiently bright blocks are selected as candidates.

{\bf Blur estimation.} This stage aims to select unblurred images with rich details. Following the same pipeline in~\cite{pech2000diatom}, we apply the Laplacian edge extraction, calculate the variance among all the output pixels and use a threshold $500$ to determine whether this image can be selected.

{\bf Color estimation.} We directly estimate the color according to~\cite{hasler2003measuring} to select images with rich color. A threshold is set to $500$ to eliminate those low-quality, gray-scale or unnatural images.

To ensure diversity, we select $97,030$ images from a total of $344,272$ images (collected from~\cite{bileschi2006streetscenes,everingham2010pascal,grubinger2006iapr,lin2014microsoft}) based on the above rules to build the dataset. We randomly select $1\%$ of them as the test set which contains $965$ images. In this paper, we use the data-balanced subset including $22,656$ images as the training set.

\subsection{Target Image Synthesis}
We propose a low-light image simulation method to synthesize realistic low-light images from normal-light images, as shown in Figure~\ref{fig_process}. This produces an adequate number of paired low/normal light images which are needed for training of learning-based methods.

\begin{figure}[t]
	\begin{center}
		\includegraphics[width=0.48\textwidth]{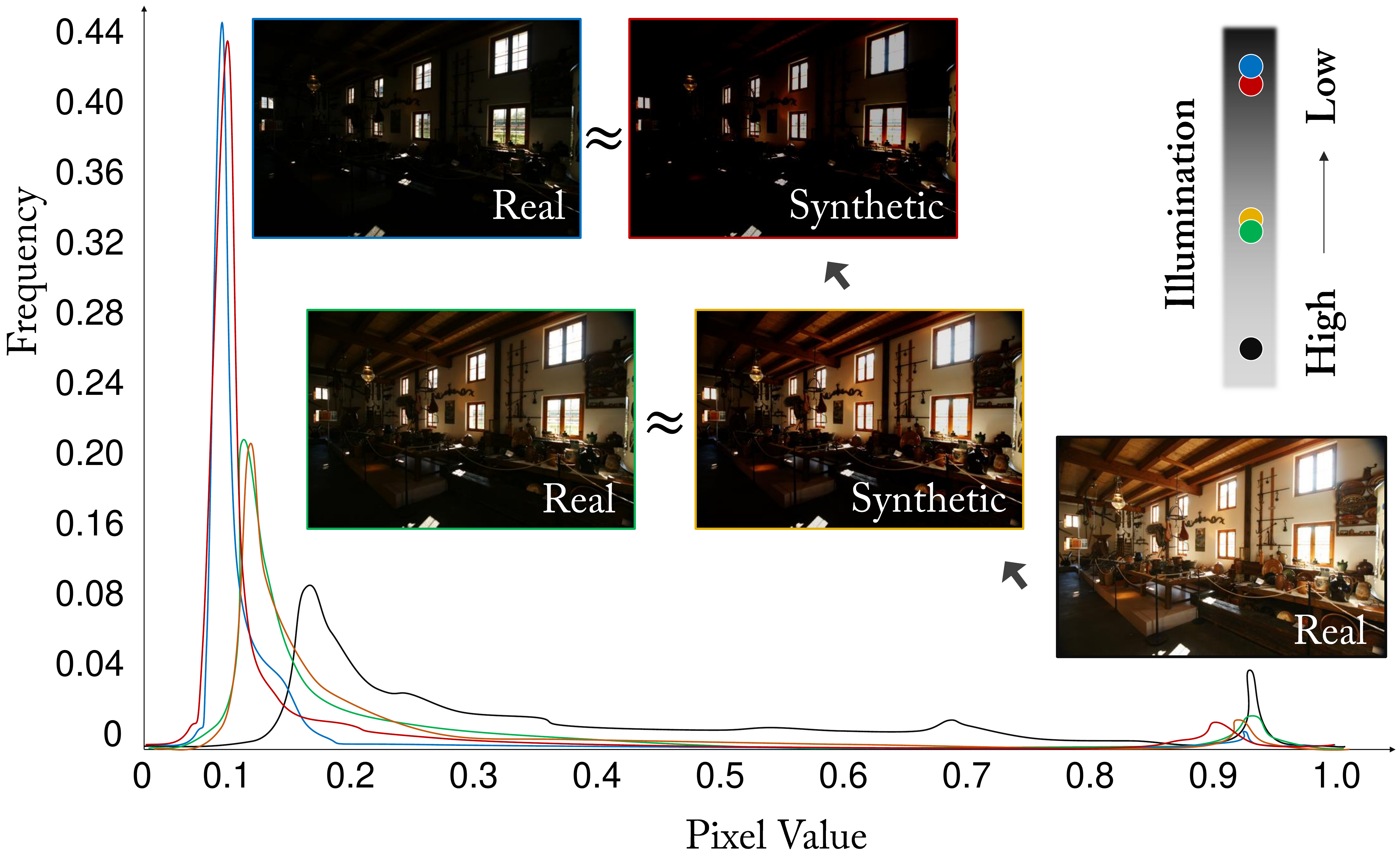} \\
	\end{center}
	\caption{Verification of the low-light simulation method: visual comparison and the histogram of \textit{Y} channel in \textit{YCbCr} between synthetic images and real different exposure images.}
	\label{fig_simulation}
\end{figure}

{\bf Low-light image synthesis.} Low-light images differ from normal images due to two dominant features: low brightness/contrast and the presence of noise. In our low-light image synthesis, we try to fit a transformation to covert the normal image to underexposed low-light image. By analyzing images with different degree of exposure, we find that the combination of linear and gamma transformation can approximate this job well. To verify this, we test on multi-exposure images and use the histogram of \textit{Y} channel in \textit{YCbCr} color space as the metric. As shown in Figure~\ref{fig_simulation}, the synthetic low-light images are approximately the same to real low-light images. The low-light image simulation pipeline (without additional noise) can be formulated as:
\begin{shrinkeq}{0ex}{
		\begin{equation}
		\begin{aligned}
		\label{lowlight_model1}
		I_{out}^{(i)} = \beta \times (\alpha \times I_{in}^{(i)})^{\gamma}, i \in \{R,G,B\}.
		\end{aligned}
		\end{equation}
}\end{shrinkeq}
where $\alpha$ and $\beta$ are linear transformations, the $X^{\gamma}$ means the gamma transformation. The three parameters is sampled from uniform distribution: $\alpha\!\sim\!U(0.9, 1), \beta\!\sim\!U(0.5, 1), \gamma\!\sim\!U(1.5, 5)$.

As for the noise, many previous methods fail to consider, while our method takes it into account. In particular, we follow~\cite{Guo2019Cbdnet,yamashita2017low} to use the Gaussian-Poisson mixed noise model and take the in-camera image processing pipeline into account to simulate real low-light noise. The noise model can be formulated as:
\begin{shrinkeq}{0ex}{
		\begin{equation}
		\begin{aligned}
		\label{noise_model1}
		I_{out} = f(M^{-1}(\mathcal{P}(M(f^{-1}(I_{in}))) + N_{G})),
		\end{aligned}
		\end{equation}
}\end{shrinkeq}
where $\mathcal{P}(x)$ represents adding Poisson noise with variance $\sigma^{2}_{p}$, $N_{G}$ is modeled as AWGN with noise variance $\sigma^{2}_{g}$, $f(x)$ stands for the camera response function, $M(x)$ is the function that convert RGB images to Bayer images and $M^{-1}(x)$ is the demosaicing function. We do not consider compression in this paper and the configuration is the same as~\cite{Guo2019Cbdnet}.

\begin{figure*}[htbp]
	\begin{center}
		\begin{overpic}[width=0.95\textwidth]{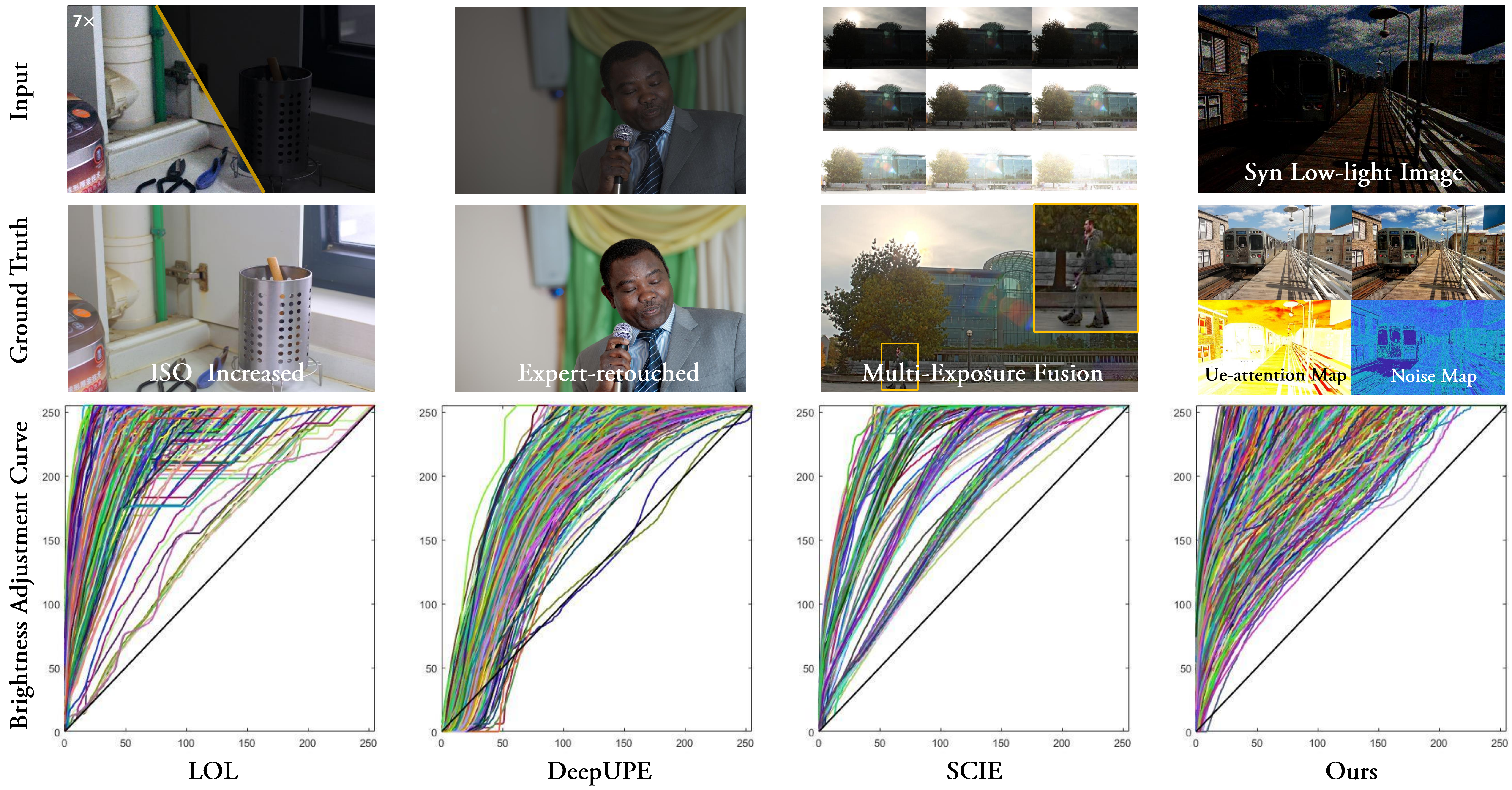}				
			\put(75,4){\bf \color{black}\scriptsize \cite{Chen2018Retinex}} 
			\put(205,4){\bf \color{black}\scriptsize \cite{wang2019underexposed}} 
			\put(315,4){\bf \color{black}\scriptsize \cite{Cai2018deep}} 
		\end{overpic}
	\end{center}
	\caption{Comparison with existing paired low-light datasets. {\bf Top:} Example images of different datasets. {\bf Bottom:} The distribution of exposure adjustment curves of different datasets.}
	\label{fig_datasetcompress}
\end{figure*}

{\bf Image contrast amplification.} The high-quality images in our dataset serve as the reference for low-light enhancement. However, directly using them to train an image-to-image regression method may result in low-contrast results (see MBLLEN~\cite{lvmbllen} results in Figure~\ref{fig_ex2}).
The possible reason caused low-contrast might that some selected images are slightly over-exposed, which will guide enhancement algorithms tend to generate slightly over-exposed results.
Besides, the smoothness caused by noise removal also result in low-contrast.

To overcome this limitation, we propose a contrast amplification method by synthesizing a new set of high-quality images as the ground truth of our second enhancement step.
In particular, we apply exposure fusion to improve the contrast/color and correct the exposure. First, we use gamma transforms to synthesize $10$ images with different exposure settings and saturation levels from each original image. Subsequently, we fuse these differently exposed images following the same routine in~\cite{mertens2007exposure} (the results called colorful images). Finally, we apply image smoothing~\cite{xu2011image} to further enhance the image details. The final output images called high-contrast images that can be used as ground truth to train a visually better low-light enhancement network.

\subsection{Comparison of Low-light Enhancement Datasets}
There are some existing datasets for low-light image enhancement. However, these datasets still have their own limitations. In this section, we highlight the differences between our synthetic dataset and other low-light image enhancement datasets, to show that our synthetic dataset is a good complement to existing datasets. The characteristics of different datasets are summarized in Table~\ref{tab_compare_dataset}.

\begin{table}[h]
	\caption{Comparison with existing low-light enhancement datasets. H(eavy), M(edium) and S(light) means the underexposed level. ``MEF" means Multi Expose Fusion methods. ``Comp." means Compatibility, which indicates whether the data acquire method can directly extend to existing public dataset for computer vision problems that have other annotations.}
	\label{tab_compare_dataset}
	\centering{\scalebox{0.90}{\scalebox{1}{\begin{tabular}{l|cccccc}
					\hline
					Dataset & Level& Source &Noise & Comp. & Scenes  \\ \hline\hline
					SID~\cite{seedark2018cvpr} & H & Camera & $\checkmark$& $\times$ & 424 \\
					LOL~\cite{Chen2018Retinex} & H-M & Camera& $\checkmark$ & $\times$ & 500  \\
					
					SICE~\cite{Cai2018deep}& H-M-S & MEF  & $\times$& $\times$ & 589  \\ 
					DeepUPE~\cite{wang2019underexposed}& M-S & Retouch & $\times$ & $\times$ & 3,000   \\
					Ours& H-M-S & Synthesis  & $\checkmark$& $\checkmark$ & {\bf 22,656} \\
					\hline
	\end{tabular}}}} \\
\end{table}

{\bf Scale and Diversity.} Having a large dataset with diverse scenes and lighting conditions is significant for training a model that can generalize well. Manually collecting images or editing images as in other datasets is a costly process, which make them hard to acquire data at scale. Therefore, existing datasets are all relatively small in size. In contrast, as our data generation is based on simulation, our method can  synthesize paired low-light and normal light images as much as needed for different scenes.

{\bf Low-light Level.} Covering a large range of low-light conditions is another important factor for the generalization capabilities of the trained models. To illustrate the range of different under-exposed levels, we calculate the exposure adjustment curve, which is the transform to the luminance channel of the low-light image (the \textit{V} component in \textit{HSV} color space) to make the luminance histogram match that of reference ground truth image. The estimated curve can serve as an estimation of the under-exposure levels, that is, steeper change of the curve indicates higher under-exposure levels.

The exposure adjustment curves for all data pair in each dataset are shown in Figure~\ref{fig_datasetcompress}. It shows that the LOL~\cite{Chen2018Retinex} dataset contains many heavy low-light images. The DeepUPE~\cite{wang2019underexposed} dataset mainly covers medium under-exposure levels. As for the SICE~\cite{Cai2018deep} dataset, the under-exposure degree is sparse, which is caused by its specific exposure pre-settings. Notice that, in this research, we use the medium exposed images as the ground truth to estimate the curves for SICE~\cite{Cai2018deep} dataset. In contrast, our synthetic dataset contains a large variety of under-exposure levels, which is useful for improving the generalization capabilities of our trained models.

{\bf Quality.} For learning-based methods, the quality of images is crucial as it directly decides the performance of training models. SCIE~\cite{Cai2018deep} uses multi-exposure image fusion result as the ground truth, which inevitably contains ghosting and blur. SID~\cite{seedark2018cvpr} prolongs exposure time to obtain high-quality night images, which may cause local overexposure and blur.
LOL~\cite{Chen2018Retinex} captures paired images by adjusting the ISO, which results in the exposure adjustment being approximately linear. In many cases, simply increasing low-light images linearly can result in good results.
As for DeepUPE~\cite{wang2019underexposed}, using retouched images as the ground truth does not have the ability to deal with noise and artifacts.
In contrast, our synthetic dataset does not have these problems. Besides, it provides the noise distribution map and exposure map that can be used as supervision to improve the performance of the trained model.

{\bf Compatibility.} Beside making the visual quality more appealing, improving the performance of other vision systems under low-light conditions is another important application for low-light enhancement. However, existing datasets do not contains manual annotations as they are only designed for visual quality enhancement. In contrast, our synthetic dataset can directly use existing public datasets (\eg, COCO~\cite{lin2014microsoft}) to render low-light images and keep their corresponding annotations, such as bounding boxes for object detection and semantic segmentation masks. Previous works ~\cite{SynRealDataFog19,SynRealDataFogECCV18} have proven that synthetic data is useful for model adaptation under adverse conditions. Thus, our synthetic dataset also has potential ability to improve the performance of fundamental vision methods to handle low-light conditions, such as object detection and semantic segmentation, etc.

In summary, our synthetic dataset has many advantages over existing datasets. Our synthetic dataset contains high quality paired pixel-aligned images with various scenes, diverse lighting conditions, and different underexposed levels. Moreover, this simulation can be applied to datasets with annotations, which is useful for model adaptation under low-light conditions. Our synthetic dataset is an important complement to existing low-light enhancement datasets.

\begin{figure*}[t]
	\begin{center}
		\includegraphics[width=1\textwidth]{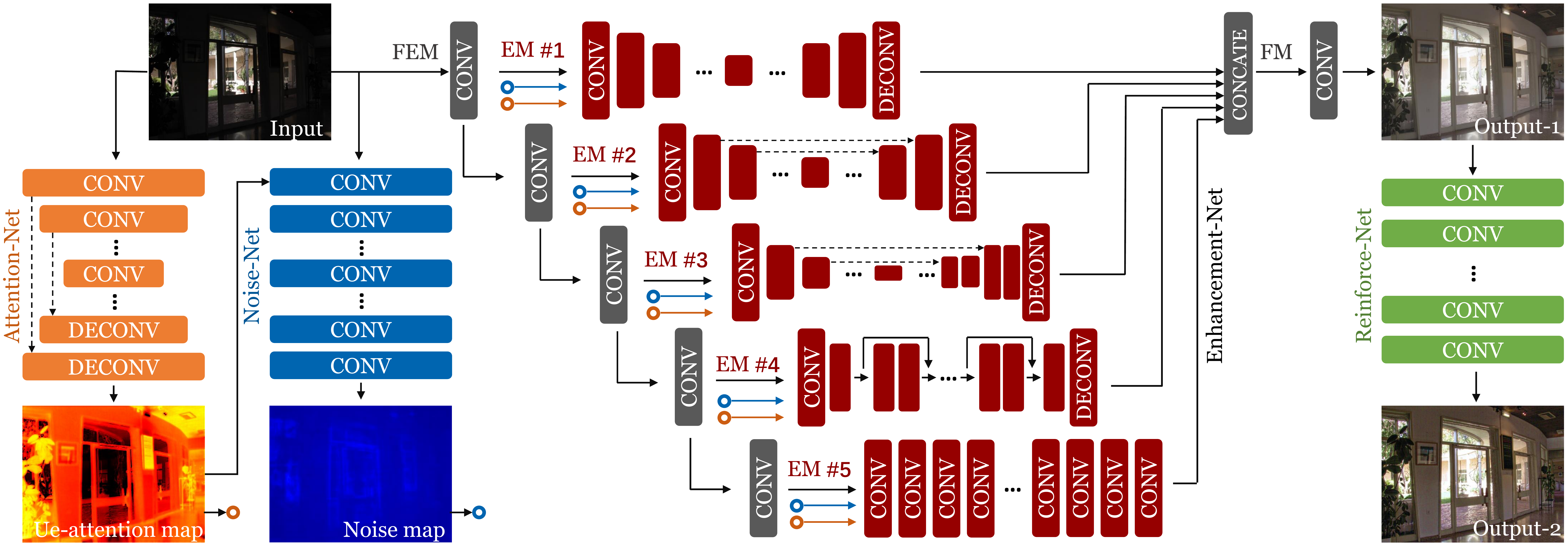}
	\end{center}
	\caption{The proposed network with four subnets. The Attention-Net and Noise-Net are used to estimate the attention of exposure and noise. The Enhancement-Net and Reinforce-Net are corresponding to the two enhancement processes. The core network is the multi-branch Enhancement-Net, which is composed of feature extraction module (FEM), enhancement module (EM) and fusion module (FM). The dashed lines represent skip connections and the circles represent discontinuous connections.}
	\label{fig_MELLEN}
\end{figure*}

\section{Attention-guided Low-light Enhancement}
In this section, we introduce the proposed attention-guided enhancement solution, including the network architecture, the loss function and the implementation details.

\subsection{Network Architecture}
We propose a fully convolutional network containing four subnets: an Attention-Net, a Noise-Net, an Enhancement Net and a Reinforce-Net. Figure~\ref{fig_MELLEN} shows the overall network architecture. The Attention-Net is designed for estimating the illumination to guide the method to pay more attention to the underexposed areas in enhancement. Similarly, the Noise-Net is designed to guide the denoising process. Under their guidance, the multi-branch Enhancement-Net can perform enhancing and denoising simultaneously. The Reinforce-Net is designed for contrast re-enhancement to solve the low-contrast limitation caused by regression. The detailed description is provided below.

\begin{figure}[t]
	\begin{center}
		\begin{overpic}[width=0.48\textwidth]{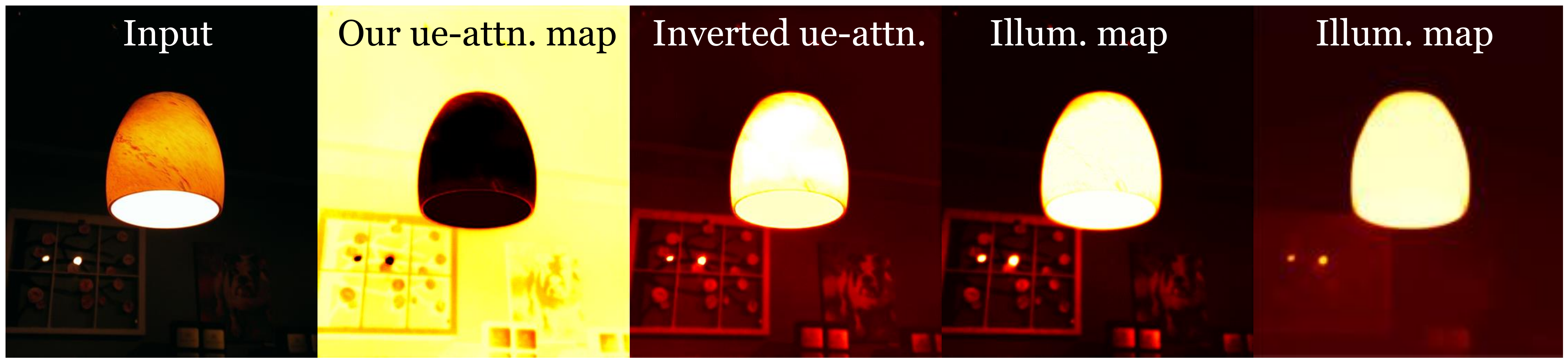}				
			\put(178,48.5){\bf \color{white}\tiny \cite{guo2017lime}} 
			\put(226,48.5){\bf \color{white}\tiny \cite{Chen2018Retinex}} 
		\end{overpic}
	\end{center}
	\caption{Comparison between our ue-attention map and the illumination maps used for retinex-based methods. Our ue-attention map can generate similar illumination information with more details.}
	\label{fig_illumination}
\end{figure}

{\bf Attention-Net.} We directly adopt U-Net in our implementation. The motivation is to provide a guidance to let Enhancement-Net correctly enhance the underexposed areas and avoid over-enhance the normally exposed areas. The output is an ue-attention map indicating the regional under-exposure level, as shown in Figure~\ref{fig_illumination}. The higher the illumination is, the lower ue-attention map values are. The ue-attention map's value range is $[0,1]$ and is determined by:
\begin{shrinkeq}{0.1ex}{
		\begin{equation}
		\begin{aligned}
		\label{attention}
		A = \frac{|max_c(I)-max_c(\mathcal{F}(I))|}{max_c(I)},
		\end{aligned}
		\end{equation}
}\end{shrinkeq}
where $max_c(x)$ returns the maximum value among three color channels, $I$ is the original bright image and $\mathcal{F}(I)$ is the synthetic under exposed image.

As shown in Figure~\ref{fig_illumination}, the inverted ue-attention map looks somewhat similar to the illumination map of the Retinex model. This infers that our ue-attention map carries important information used by the popular Retinex model. On the other hand, using our inverted ue-attention map in Retinex model still cannot ensure satisfactory results. This is because the Retinex-based solution faces difficulties in handling black regions (see black regions in Figure~\ref{fig_first}) and will result in noise amplification (see LIME results in Figure~\ref{fig_nature_compare}). Therefore, we propose to use the ue-attention map as a guidance for our Enhancement Net introduced later.

{\bf Noise-Net.} The image noise can be easily confused with image textures, causing unwanted blurring effect after applying simple denoising methods. Estimating the noise distribution beforehand and making the denoising adaptive may help reduce such an effect. The noise map's value range is $[0,1]$ and is determined by:
\begin{shrinkeq}{0.1ex}{
		\begin{equation}
		\begin{aligned}
		\label{noise}
		N = max_c(\frac{|\mathcal{F}_n(I)-\mathcal{F}(I)|}{\mathcal{F}(I)}),
		\end{aligned}
		\end{equation}
}\end{shrinkeq}
where $max_c(x)$ returns the maximum value among three color channels, $\mathcal{F}_n(I)$ is the synthetic low-light image and $\mathcal{F}(I)$ is the synthetic under exposed image.

Note that the noise distribution is highly related to the distribution of exposure, and thus we propose to use the ue-attention map to help derive a noise map. Under their guidance, the enhancement-net can perform denoising effectively. The Noise-Net is composed of dilated convolutional layers to increase the receptive field, which is conducive to noise estimation.

{\bf Enhancement-Net.} The motivation is to decompose the enhancement problem into several sub-problems of different aspects (such as noise removal, texture preserving, color correction and so on) and solve them respectively to produce the final output via multi-branch fusion. It is the core component of the proposed network and it consists of three types of modules: the feature extraction module (FEM), the enhancement module (EM) and the fusion module (FM). {\bf FEM} is a single stream network with several convolutional layers, each of which uses $3\times3$ kernels, stride of $1$ and ReLU nonlinearity. The output of each layer is both the input to the next layer and also the input to the corresponding subnet of EM. {\bf EMs} are modules following each convolutional layer of the FEM. The input to EM is the output of a certain layer in FEM, and the output size is the same as the input. {\bf FM} accepts the outputs of all EMs to produce the final enhanced image. We concatenate all the outputs from EMs in the color channel dimension and use the $1\times1$ convolution kernel to merge them, which equals to the weighted summation with learnable weights.

We propose five different EM structures. As shown in Figure~\ref{fig_MELLEN}, the design of EM follows U-Net~\cite{ronneberger2015u} and Res-Net~\cite{he2016deep} which have been proven effective extensively. In brief, EM-1 is a stack of convolutional and deconvolutional layers with large kernel size. EM-2 and EM-3 has U-Net like structures, and the difference is the skip connection realization and the feature map size. EM-4 has a Res-Net like structure. We remove the Batch-Normalization~\cite{ioffe2015batch} and use just a few res-blocks to reduce the model parameter. EM-5 is composed of dilated convolutional layers whose output size is the same as the input.

{\bf Reinforce-Net.} The motivation is to overcome the low-contrast drawback and improve the details (see the difference between MBLLEN~\cite{lvmbllen} and ours in Figure~\ref{fig_ex2}). Previous research~\cite{Qifeng2017Fast} demonstrates the effectiveness of dilated convolution in image processing. Therefore, we use a similar network to improve contrast and details simultaneously.

\subsection{Loss Function}
In order to improve the image quality both qualitatively and quantitatively, we propose a new loss function by further considering the structural information, perceptual information and regional difference of the image. It is expressed as:
\begin{shrinkeq}{0.5ex}{
		\begin{equation}
		\begin{aligned}
		\label{align_loss}
		\mathcal{L} = \omega_a\mathcal{L}_a + \omega_n\mathcal{L}_n + \omega_e\mathcal{L}_e + \omega_r\mathcal{L}_r,
		\end{aligned}
		\end{equation}
}\end{shrinkeq}
where the $\mathcal{L}_a$, $\mathcal{L}_n$, $\mathcal{L}_e$ and $\mathcal{L}_r$ represent the loss function of Attention-Net, Noise-Net, Enhancement-Net and Re-inforce-Net, and $\omega_a,\omega_n,\omega_e,\omega_r$ are the corresponding coefficients. The details of the four loss functions are given below.

{\bf Attention-Net loss.} To obtain the correct ue-atte-ntion map for guiding the Enhancement-Net, we use the $L2$ error metric to measure the prediction error as:
\begin{shrinkeq}{0.5ex}{
		\begin{equation}
		\begin{aligned}
		\label{Attention_loss}
		\mathcal{L}_a = \| \mathcal{F}_{a}(I)  - A \|^2,
		\end{aligned}
		\end{equation}
}\end{shrinkeq}
where $I$ is the input image, $\mathcal{F}_{a}(I)$ and $A$ are the predicted and expected ue-attention maps.

{\bf Noise-Net loss.} Similarly, we use the $L1$ error metric to measure the prediction error of the Noise-Net as:
\begin{shrinkeq}{0.5ex}{
		\begin{equation}
		\begin{aligned}
		\label{noise_loss}
		\mathcal{L}_n = \| \mathcal{F}_{n}(I, A')  - N \|^1,
		\end{aligned}
		\end{equation}
}\end{shrinkeq}
where $A'=\mathcal{F}_{a}(I)$, $\mathcal{F}_{n}(I, A')$ and $N$ are the predicted and expected noise maps.

{\bf Enhancement-Net loss.} Due to the low brightness of the image, only using common error metrics such as \textit{mse} or \textit{mae} may cause structure distortion such as blur effect and artifacts. We design a new loss that consists of four components to improve the visual quality. It is defined as:
\begin{shrinkeq}{0.5ex}{
		\begin{equation}
		\begin{aligned}
		\label{enhancement_loss}
		\mathcal{L}_e = \omega_{eb}\mathcal{L}_{eb} + \omega_{es}\mathcal{L}_{es} + \omega_{ep}\mathcal{L}_{ep} + \omega_{er}\mathcal{L}_{er},
		\end{aligned}
		\end{equation}
}\end{shrinkeq}
where the $\mathcal{L}_{eb}$, $\mathcal{L}_{es}$, $\mathcal{L}_{ep}$ and $\mathcal{L}_{er}$ represent bright loss, structural loss, perceptual loss and regional loss. And $\omega_{eb}$, $\omega_{es}$, $\omega_{ep}$ and $\omega_{er}$ are the corresponding coefficients.

The bright loss is designed to ensure that the enhanced results have sufficient brightness. It is defined as:
\begin{shrinkeq}{0.5ex}{
		\begin{equation}
		\begin{aligned}
		\label{bright_loss}
		\mathcal{L}_{eb} = \|\mathcal{S}(\mathcal{F}_{e}(I, A', N') -\widetilde{I})\|^1, 
		\end{aligned}
		\end{equation}
}\end{shrinkeq}
where $\mathcal{F}_{e}(I, A', N')$ and $\widetilde{I}$ are the predicted and expected enhancement images. $\mathcal{S}$ is defined as: $\mathcal{S}(x<0)=-\lambda x, \mathcal{S}(x\ge0)=x, s.t.~ \lambda>1$.

The structural loss is introduced to preserve the image structure and avoid blurring. We use the well-known image quality assessment algorithm SSIM~\cite{wang2004image} to build our structure loss.
The structural loss is defined as:
\begin{shrinkeq}{0.5ex}{
		\begin{align}
		\label{align_ssim}
		\mathcal{L}_{es} = 1 - \frac{1}{N}\sum_{p\in img}\frac{2\mu_x \mu_y + C_1}{\mu_x^2+\mu_y^2+C_1}\cdot \frac{2\sigma_{xy} + C_2}{\sigma_x^2+\sigma_y^2+C_2},
		\end{align}
}\end{shrinkeq}
where $\mu_x$ and $\mu_y$ are pixel value averages, $\sigma_x^2$ and $\sigma_y^2$ are variances, $\sigma_{xy}$ is the covariance, and $C_1$ and $C_2$ are constants to prevent the denominator to zero.

The perceptual loss is introduced to use higher-level information to improve the visual quality. We use the well-behaved VGG network~\cite{simonyan2014very} as the content extractor~\cite{ledig2016photo}. In particular, we define the perceptual loss based on the output of the ReLU activation layers of the pre-trained VGG-19 network.	The perceptual loss is defined as follows:
\begin{shrinkeq}{0.5ex}{
		\begin{equation}
		\begin{aligned}
		\label{align_vggloss}
		\mathcal{L}_{ep} = \frac{1}{w_{ij}h_{ij}c_{ij}}\sum_{x=1}^{w_{ij}}\sum_{y=1}^{h_{ij}}\sum_{z=1}^{c_{ij}}
		\lVert\phi_{ij}(I')_{xyz}-\phi_{ij}(\widetilde{I})_{xyz}\lVert,
		\end{aligned}
		\end{equation}
}\end{shrinkeq}
where $I'=\mathcal{F}_{e}(I, A', N')$ and $\widetilde{I}$ are the predicted and expected enhancement images, and $w_{ij}$ , $h_{ij}$ and $c_{ij}$ describe the dimensions of the respective feature maps within the VGG-19 network. Besides, $\phi_{ij}$ indicates the feature map obtained by $j$-th convolution layer in $i$-th block of the VGG-19 Network.

For low-light image enhancement, except taking the image as a whole, we should pay more attention to the underexposed regions. We propose the regional loss to balances the degree of enhancement for different regions. It is defined as:

\begin{shrinkeq}{0.5ex}{
		\begin{small}
			\begin{equation}
			\begin{aligned}
			\label{align_regionloss}
			\mathcal{L}_{er} = \| I'\cdot A'  - \widetilde{I}\cdot A' \|^1 + 1 - ssim(I'\cdot A', \widetilde{I}\cdot A' )
			\end{aligned}
			\end{equation}
		\end{small}
}\end{shrinkeq}
where $ssim(\cdot)$ represents the image quality assessment algorithm SSIM~\cite{wang2004image} and $A'$ is the predicted ue-attention map which is used as the guidance.

{\bf Reinforce-Net loss.} Similar to the Enhancement-Net loss, the Reinforce-Net loss is defined as:
\begin{shrinkeq}{0.5ex}{
		\begin{equation}
		\begin{aligned}
		\label{Reinforce_loss}
		\mathcal{L}_r = \omega_{rb}\mathcal{L}_{rb} + \omega_{rs}\mathcal{L}_{rs} + \omega_{rp}\mathcal{L}_{rp},
		\end{aligned}
		\end{equation}
}\end{shrinkeq}
where $\mathcal{L}_{rb}$, $\mathcal{L}_{rs}$ and $\mathcal{L}_{rp}$ represent bright loss, structural loss and perceptual loss, and are the same as $\mathcal{L}_{rb}$, $\mathcal{L}_{rs}$ and $\mathcal{L}_{rp}$.
In the experiments, we empirically set $\!\lambda\!=\!10$, $\omega_{a},\omega_{n},\omega_{e},\omega_{r}\!=\!\{100,10,10,1\}$, $\omega_{eb},\omega_{es},\omega_{ep},\omega_{er}\!=\!\{1,1,0.35,5\}$, $\omega_{rb},\omega_{rs},\omega_{rp}\!=\!\{1,1,0.35\}$.

\subsection{Implementation Details}
Our implementation is done with Keras~\cite{chollet2015keras} and Tensorflow ~\cite{abadi2016tensorflow}. The proposed network can be quickly converged after being trained for $20$ epochs on a Titan-X GPU using the proposed dataset. In order to prevent overfitting, we use random clipping, flipping and rotating for data augmentation. We set the batch-size to $8$ and the size of random clipping patches to $256\times256\times3$. The input image values is scaled to $[0, 1]$. We use the output of the fourth convolutional layer in the third block of VGG-19 network as the perceptual loss extraction layer.

In the experiment, training is done using the Adam optimizer~\cite{kingma2014adam} with parameters of $\alpha = 0.0002$, $\beta_{1} = 0.9$, $\beta_{2} = 0.999$ and $\epsilon = 10^{-8}$. We also use the learning rate decay strategy, which reduces the learning rate to $98\%$ before the next epoch. At the same time, we reduce the learning rate to $50\%$ when the loss metric has stopped improving.

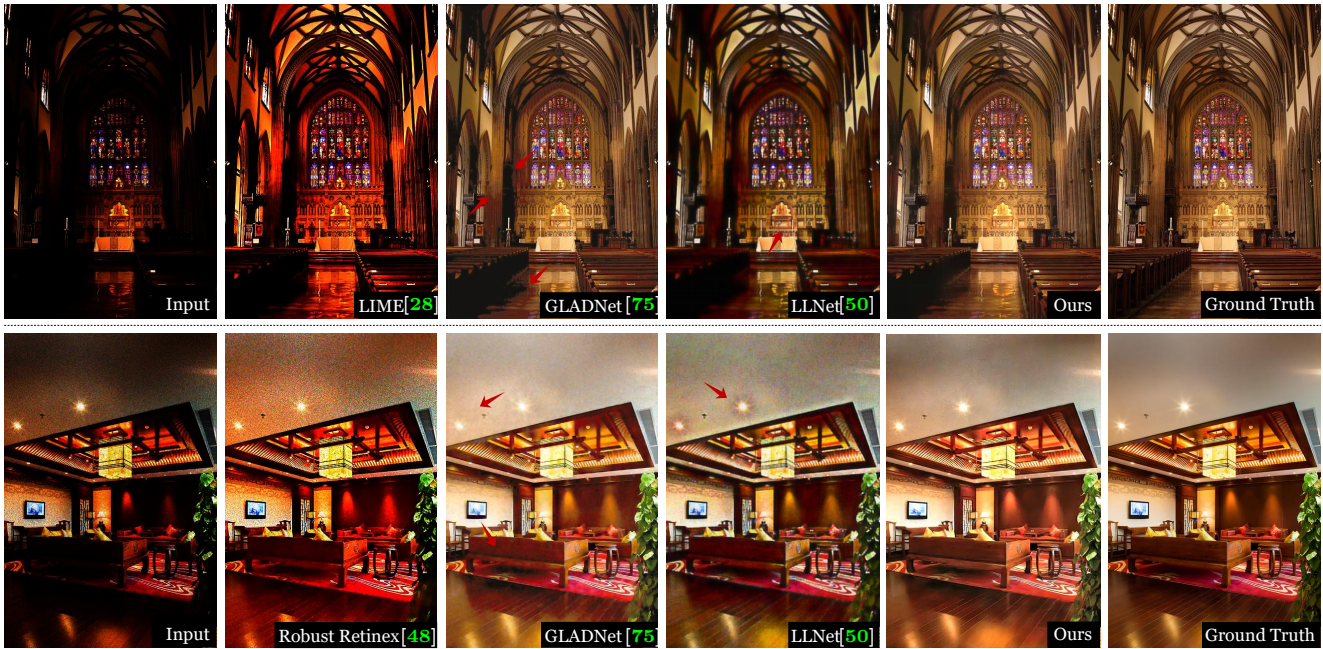
\begin{figure*}[htbp]
	\begin{center}
		\begin{overpic}[width=1\textwidth]{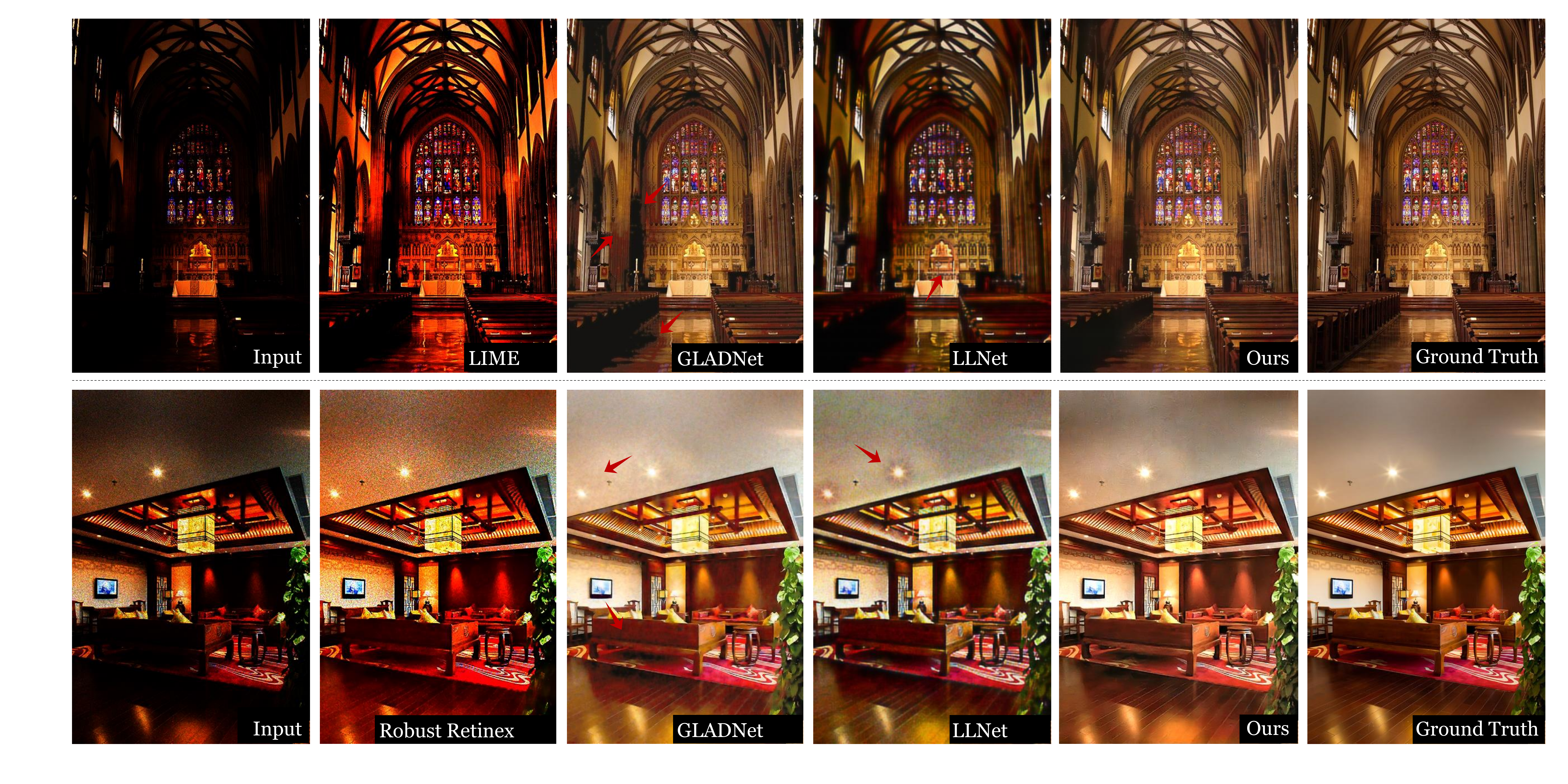}
			\put(150,128){\bf \color{white}\scriptsize \cite{guo2017lime}} 
			\put(233,128){\bf \color{white}\scriptsize \cite{wang2018gladnet}} 
			\put(313,128){\bf \color{white}\scriptsize \cite{lore2017llnet}} 
			
			\put(149,3){\bf \color{white}\scriptsize \cite{li2018structure}} 
			\put(233,3){\bf \color{white}\scriptsize \cite{wang2018gladnet}} 
			\put(313,3){\bf \color{white}\scriptsize \cite{lore2017llnet}} 
		\end{overpic}
	\end{center}
	\caption{Visual comparison on synthetic low-light images. We fine tune the GLADNet~\cite{wang2018gladnet} using our synthetic datasets. Please zoom in for a better view.}
	\label{fig_ex1}
\end{figure*}

\section{Experimental Evaluation}
We compare our method with existing methods through extensive experiments. We use the publicly-available codes with recommended parameter settings. In quantitative comparison, we used PSNR and SSIM~\cite{wang2004image}, along with some recently proposed metrics \textit{Average Brightness} (AB)~\cite{chen2006gray}, \textit{Visual Information Fidelity} (VIF)~\cite{sheikh2006image}, \textit{Lightness Order Error} (LOE)~\cite{ying2017bio}, \textit{Tone Mapped Image Quality Index} (TMQI)~\cite{yeganeh2013objective} and \textit{Learned Perceptual Image Patch Similarity Metric} (LPIPS)\cite{zhang2018perceptual}.
For all metrics higher number means better, except LPIPS, LOE and AB. Note that in the tables below, {\color{red}{red}}, {\color{lvgreen}{green}} and {\color{lvblue}{blue}} colors indicate the best, second, and third place results, respectively.

\begin{table}[tb]
	\caption{Quantitative comparison of synthetic low-light image (without additional noise) enhancement. ``$\uparrow$" indicates the higher the better, ``$\downarrow$" indicates the lower the better, `$`\Downarrow$" indicates the lower absolute value the better. }
	\scalebox{0.70}{\begin{tabular}{l|rrrrrrr}
			\hline
			~         & PSNR$\uparrow$ & SSIM$\uparrow$ & LPIPS$\downarrow$ & VIF$\uparrow$ & LOE$\downarrow$ & TMQI$\uparrow$ & AB$\Downarrow$ \\ \hline\hline
			Input & 11.99 & 0.45 & 0.26& 0.33 & 677.85 & 0.80 & -59.22\\
			BIMEF~\cite{ying2017bio} & 18.28 & 0.76 & \color{lvblue}{\bf{0.11}}& 0.49 & 550.20 & 0.85 & -28.06\\
			LIME~\cite{guo2017lime} & 15.80 & 0.68 & 0.20& 0.48 & 1121.17 & 0.80 &  \color{lvgreen}{\bf{-2.46}}\\
			MSRCR~\cite{jobson1997multiscale} & 14.87 & 0.72 & 0.15 & 0.52 & 1249.24 & 0.82 & 35.07\\
			MF~\cite{fu2016mf} & 15.89 & 0.68 & 0.18& 0.44 & 766.00 & 0.83 & -36.88\\
			SRIE~\cite{fu2016srie} & 13.83 & 0.56 & 0.21& 0.37 & 787.42 & 0.82 & -47.86\\
			Dong~\cite{dong2011fast} & 15.37 & 0.65 & 0.22& 0.35 & 1228.49 & 0.81 & -33.80\\
			NPE~\cite{wang2013naturalness} & 14.93 & 0.66 & 0.18& 0.42 & 875.15 & 0.83 & -41.35\\
			DHECI~\cite{nakai2013dheci} & 18.13 & 0.76 & 0.17& 0.39 &  547.12 & 0.87 & -17.37\\
			BPDHE~\cite{ibrahim2007bpdhe} & 13.62 & 0.60 & 0.24& 0.34 & 609.89 & 0.82 & -47.82\\
			HE & 17.88 & 0.76 & 0.18& 0.47 & 596.67 &  \color{lvblue}{\bf{0.88}} & 19.24\\
			Ying~\cite{ying2017newiccv} &  19.21 &  \color{lvblue}{\bf{0.80}} & \color{lvblue}{\bf{0.11}}&  0.56 & 778.67 & 0.83 & -9.28\\
			WAHE~\cite{arici2009wahe} & 15.46 & 0.65 & 0.18& 0.44 & 564.83 & 0.84 & -39.38\\
			JED~\cite{ren2018joint} & 16.11 & 0.65 & 0.21 & 0.41 & 1212.66 & 0.82 & -25.95 \\
			Robust~\cite{li2018structure} &16.83 & 0.69 & 0.20 & 0.47 & 1052.22 & 0.82 & -22.09\\
			LLNet~\cite{lore2017llnet} & 20.11 &  \color{lvblue}{\bf{0.80}} & 0.39& 0.40 & 1088.43 & 0.87 & 4.30\\
			DeepUPE~\cite{wang2019underexposed} & 16.55 & 0.64 & 0.17 & 0.55 & 516.47 & 0.84 & -30.48 \\
			GLADNet~\cite{wang2018gladnet} & \color{lvgreen}{\bf{24.57}} & \color{lvgreen}{\bf{0.90}} & \color{lvgreen}{\bf{0.09}} & \color{lvblue}{\bf{0.62}} & \color{lvgreen}{\bf{513.18}} & \color{lvgreen}{\bf{0.91}} & 5.52\\
			MBLLEN~\cite{lvmbllen} &  \color{lvblue}{\bf{24.21}} &  \color{lvgreen}{\bf{0.90}} & \color{red}{\bf{0.08}}& \color{lvgreen}{\bf{0.63}} & \color{lvblue}{\bf{536.75}} & \color{lvgreen}{\bf{0.91}} &  \color{lvblue}{\bf{-3.66}}\\
			Ours &  \color{red}{\bf{25.24}} &  \color{red}{\bf{0.94}} & \color{red}{\bf{0.08}}&  \color{red}{\bf{0.67}} &  \color{red}{\bf{495.48}} &  \color{red}{\bf{0.93}} &  \color{red}{\bf{2.04}}\\
			\hline
	\end{tabular}}\\
	\label{tab_darkenhance}
\end{table}

\begin{table}[tb]
	\caption{Quantitative comparison of synthetic low-light images (with additional noise) enhancement. ``$\uparrow$" indicates the higher the better, ``$\downarrow$" indicates the lower the better, `$`\Downarrow$" indicates the lower absolute value the better.}
	\label{tab_LL}
	\scalebox{0.70}{\begin{tabular}{l|rrrrrrr}
			\hline
			~ & PSNR$\uparrow$ & SSIM$\uparrow$ & LPIPS$\downarrow$ & VIF$\uparrow$ & LOE$\downarrow$ & TMQI$\uparrow$ & AB$\Downarrow$\\
			\hline\hline
			Input & 11.23 & 0.37 & 0.41 & 0.23 &  925.06 & 0.77 & -65.32\\
			BIMEF~\cite{ying2017bio} & 16.57 & 0.64 & 0.32 &  \color{lvblue}{\bf{0.28}} & 978.96 & 0.83 & -32.65\\
			LIME~\cite{guo2017lime} & 14.79 & 0.59 & 0.34 & 0.26 & 1462.64 & 0.79 &  -7.39\\
			MSRCR~\cite{jobson1997multiscale} & 14.83 & 0.62 & 0.34 & 0.27 & 1559.05 & 0.84 & 30.98\\
			MF~\cite{fu2016mf} & 15.29 & 0.59 & 0.33 & 0.26 & 1095.33 & 0.82 & -37.46\\
			SRIE~\cite{fu2016srie} & 13.10 & 0.48 & 0.37& 0.25 & 1095.30 & 0.80 & -52.53\\
			Dong~\cite{dong2011fast} & 14.69 & 0.56 & 0.35& 0.21 & 1592.27 & 0.79 & -33.99\\
			NPE~\cite{wang2013naturalness} & 14.56 & 0.58 & 0.33& 0.25 & 1302.10 & 0.82 & -41.17\\
			DHECI~\cite{nakai2013dheci} & 16.57 & 0.61 & 0.37& 0.23 &  924.78 & 0.86 & -15.20\\
			BPDHE~\cite{ibrahim2007bpdhe} & 12.60 & 0.48 & 0.38& 0.23 & 925.56 & 0.79 & -54.66\\
			HE & 16.65 & 0.64 & 0.36& 0.26 & 1036.22 &  0.87 & 20.21\\
			Ying~\cite{ying2017newiccv} &  17.18 &  0.67 & 0.31&  \color{lvblue}{\bf{0.28}} & 1152.94 & 0.83 &  -13.97\\
			WAHE~\cite{arici2009wahe} & 13.97 & 0.52 &0.36 & 0.27 & 935.21 & 0.81 & -46.87\\
			JED~\cite{ren2018joint} & 13.70 & 0.48 & 0.46 & 0.22 & 1531.84 & 0.77 & -33.11\\
			Robust~\cite{li2018structure} & 14.03 & 0.50 & 0.46 & 0.23 & 1448.03 & 0.77 & -29.09 \\
			LLNet~\cite{lore2017llnet} & 18.40 & 0.69 & 0.56& 0.26 & 1168.75 & 0.85 & -5.25\\
			DeepUPE~\cite{wang2019underexposed} &14.94 & 0.53 & 0.35 & 0.25 & 1084.08 & 0.81 & -36.53\\
			GLADNet~\cite{wang2018gladnet} & \color{lvgreen}{\bf{19.86}} & \color{lvgreen}{\bf{0.76}} & \color{lvgreen}{\bf{0.19}} & \color{lvgreen}{\bf{0.30}} & \color{lvgreen}{\bf{796.87}} & \color{lvblue}{\bf{0.88}} &  \color{lvblue}{\bf{5.09}} \\
			MBLLEN~\cite{lvmbllen} & \color{lvblue}{\bf{19.27}} &  \color{lvblue}{\bf{0.73}} & \color{lvblue}{\bf{0.23}}& \color{lvgreen}{\bf{0.30}} & \color{lvblue}{\bf{864.57}} & \color{lvgreen}{\bf{0.89}} &  \color{lvgreen}{\bf{-4.87}}\\
			Ours &  \color{red}{\bf{20.84}} &  \color{red}{\bf{0.82}}& \color{red}{\bf{0.17}} &  \color{red}{\bf{0.33}} &  \color{red}{\bf{785.64}} &  \color{red}{\bf{0.91}} & \color{red}{\bf{4.36}}
			\\ \hline
	\end{tabular}}\\
	\centering
\end{table}


Our experiment is organized as following. First, we make qualitative and quantitative comparisons based on our synthetic dataset and two public-available real low-light datasets. Second, we make visual comparisons with state-of-the-art methods on natural low-light images and provide a user study. We also show the robustness of our method and the benefit to some high-level tasks. Finally, we provide an ablation study to evaluate the effect of different elements and discuss unsatisfying cases.

\subsection{Experiments on Synthetic Datasets}
{\bf Direct comparison}. We compare our method with state-of-the-art methods on our synthetic dataset. Since most methods do not have the ability to remove noise, we combine them with the state-of-the-art denoising method CBDNet~\cite{Guo2019Cbdnet} to produce the final comparison results. We fine tune the GLADNet~\cite{wang2018gladnet} and LLNet~\cite{lore2017llnet} for fair comparison. Quantitative comparison results are shown in Table~\ref{tab_darkenhance} and Table~\ref{tab_LL}. Our result significantly outperforms other methods in all quality metrics, which fully demonstrates the superiority of our approach.

\begin{figure}[t]
	\begin{center}
		\includegraphics[width=0.48\textwidth]{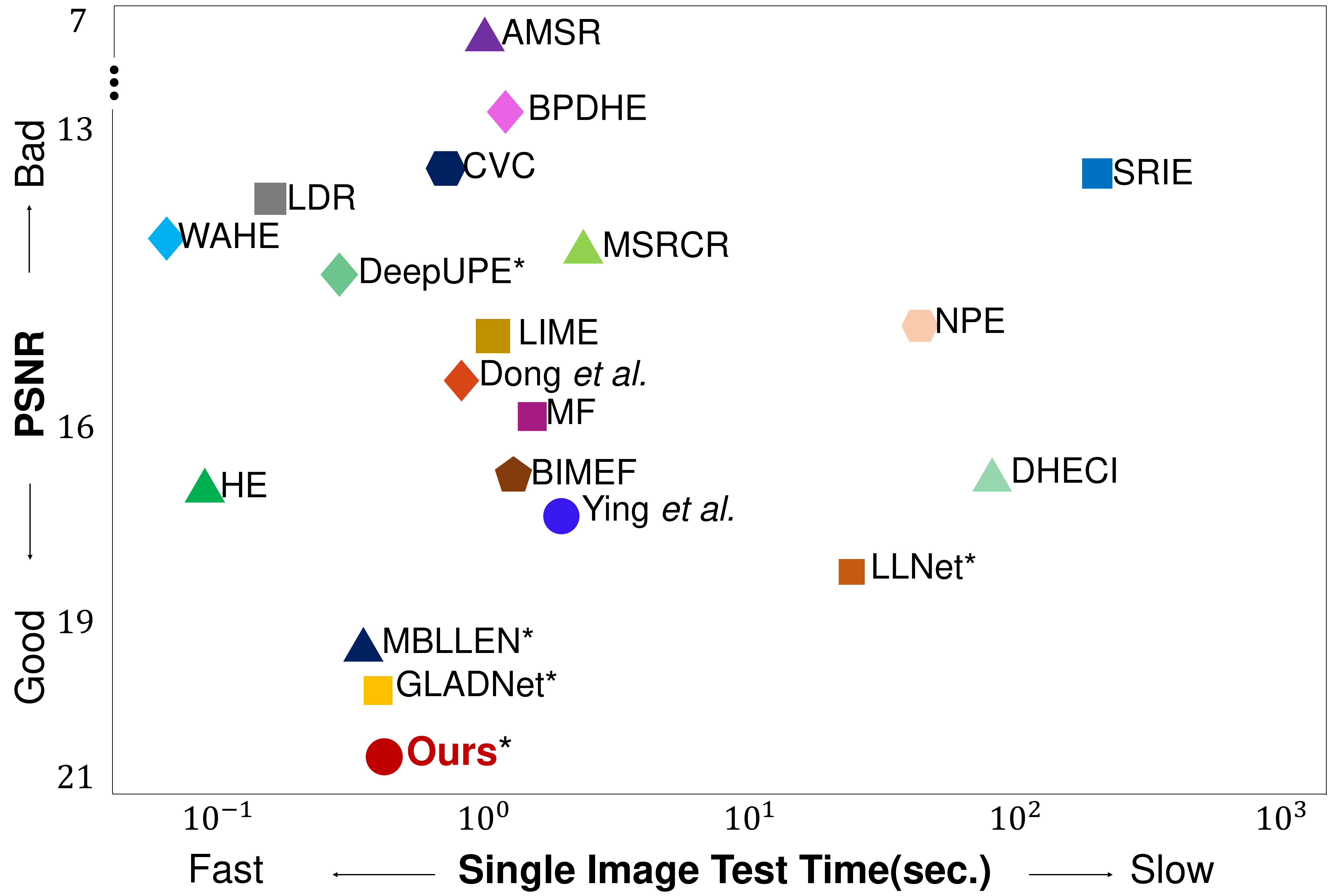}
	\end{center}
	\caption{Runtime and performance comparison of different enhancement methods. Test machine is a PC with Intel i5-8400 CPU, $16$ GB memory and NVIDIA Titan-Xp GPU. ``*" represents using GPU.}
	\label{runtimefig}
\end{figure}

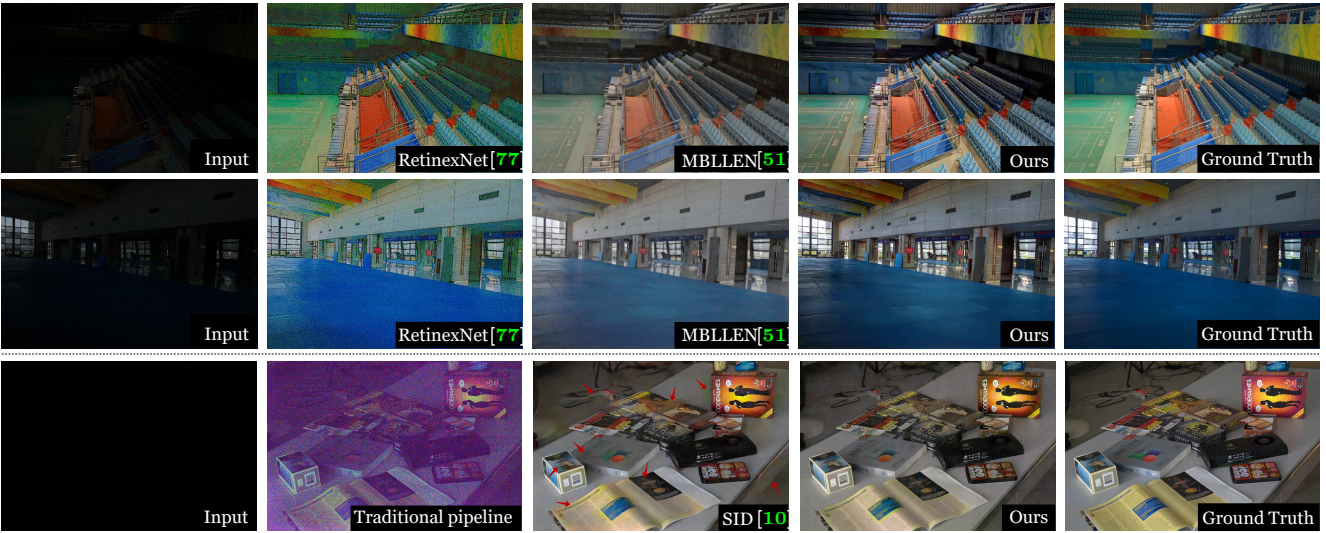
\begin{figure*}[t]
	\begin{center}
		\begin{overpic}[width=1\textwidth]{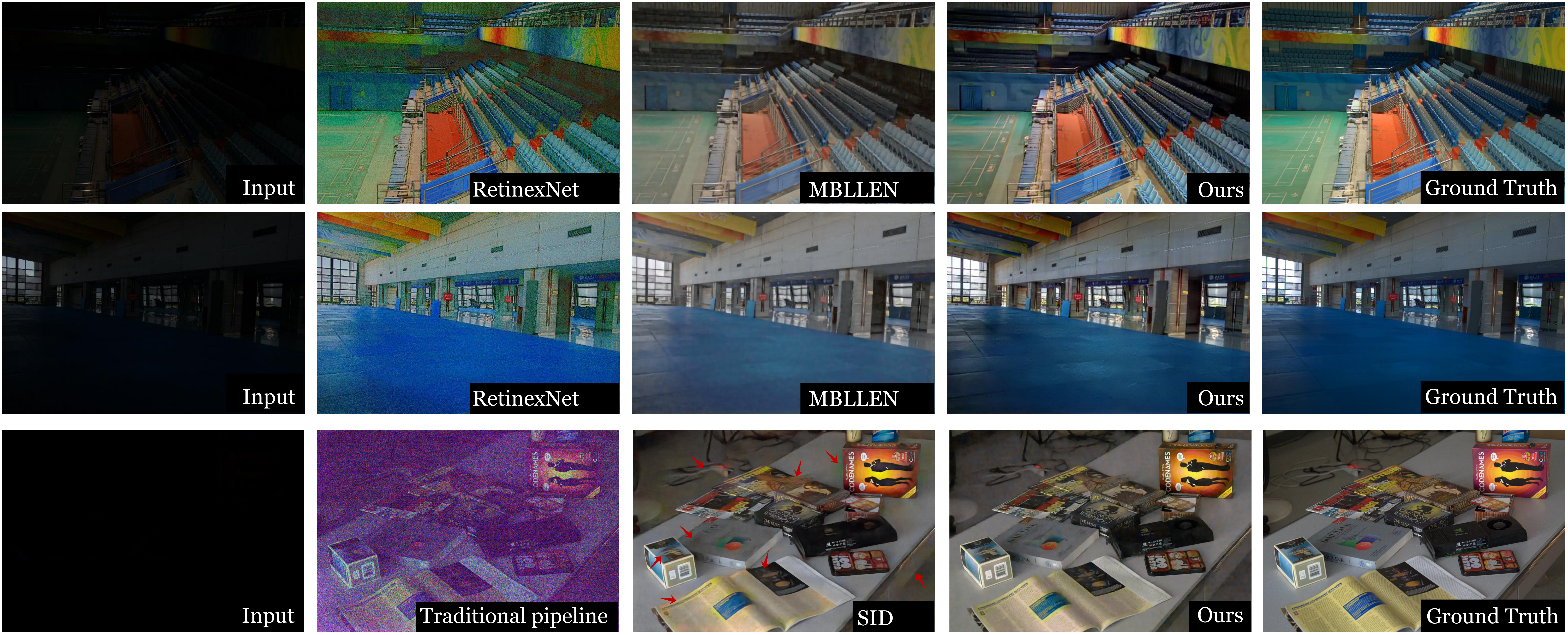}
			\put(283,3.5){\bf \color{white}\scriptsize \cite{seedark2018cvpr}} 
			\put(183,72.5){\bf \color{white}\scriptsize \cite{Chen2018Retinex}} 
			\put(283,72.5){\bf \color{white}\scriptsize \cite{lvmbllen}} 
			\put(183,139){\bf \color{white}\scriptsize \cite{Chen2018Retinex}} 
			\put(283,139){\bf \color{white}\scriptsize \cite{lvmbllen}} 
		\end{overpic}
	\end{center}
	\caption{Visual comparison on the LOL dataset (row 1 and 2) and the SID dataset (row 3). Please zoom in for a better view.}
	\label{fig_ex2}
\end{figure*}

Representative results are visually shown in Figure~\ref{fig_ex1}. By checking the details, it is clear that our method achieves better visual effects, including good brightness/contrast and less artifacts. Please zoom in to compare the details.


{\bf Efficiency comparison}. In addition to the result quality, efficiency is also an important metric to algorithms. In order to demonstrate the superiority of our method, we use $10$ HD images with size $1920\times1080$ as the benchmark to test running time. In order to more intuitively demonstrate the relationship between performance and efficiency, we show Figure~\ref{runtimefig}. Our method performs well in terms of both quality and efficiency. Notice that, JED~\cite{ren2018joint} and Robust~\cite{li2018structure} need large computational resources, which will cause out-of-memory problem when processing large images. Due to the MLP architecture, LLNet~\cite{lore2017llnet} needs to enhance large images one patch by one patch, which will limits its efficiency.


\subsection{Experiments on Real Datasets}
Besides synthetic datasets, our method also performs well on real low-light image datasets. We evaluate the performance based on two public-available real low-light datasets and show the visual comparison on challenging images.

{\bf LOL dataset.} This dataset is captured by control the exposure and ISO in the daytime. We fine-tune our model using this dataset to compare with RetinexNet~\cite{seedark2018cvpr}, which is trained on the LOL dataset. In addition, we replace the Enhancement-Net by a standard U-Net to build a lightweight version. Following PPCN~\cite{Hui-PPCN-2018}, we also adopt knowledge transfer to further promote its performance. Quantitative comparison is shown in Table~\ref{tab_ex2_1}. For both quality and efficiency comparisons, our method performs better, manifesting that our method effectively learns the adjustment and restoration. Visual comparison is shown in Figure~\ref{fig_ex2}. Compared with RetinexNet~\cite{Chen2018Retinex} and MBLLEN~\cite{lvmbllen}, our results with clear details, better contrast, normal brightness and natural white balance.

{\bf SID dataset.} This dataset contains raw short-expo-sure images with corresponding long-exposure reference images and is benchmarking single-image processing of extremely low-light raw images. Due to the larger bit depth, raw images are more suitable for extremely low-light scenes compared with rgb images. Different from traditional pipelines, SID~\cite{seedark2018cvpr} develop a pipeline based on an end-to-end network and achieve excellent results. Need to notice that, processing low-light raw images is a related but not identical problem. However, to prove the ability of our multi-branch network, we use the same configuration except that the network is replaced by our Enhancement-Net. Quantitative comparison is shown in Table~\ref{tab_ex2_1}. Our model is lightweight and more efficient, but achieves comparable enhancement quality. In addition, our results have better visual effects as shown in Figure~\ref{fig_ex2}.

\subsection{Experiments on Real Images}
In this section, we evaluate our method on real low-light images, including natural, monochrome and game scenes. We also show the benefit to object detection and semantic segmentation under low-light environment by directly using our method as the pre-processing.

\begin{table}[t]
	\caption{Quantitative comparison between our method and state-of-the-arts on the LOL dataset and the SID dataset. ``ours-1" means the result of the Enhancement-Net, ``ours-2" means the result of the Reinforce-Net.}
	\label{tab_ex2_1}
	\centering{\scalebox{0.81}{\scalebox{1}{\begin{tabular}{l|ccccc}
					\hline
					Method  & PSNR& SSIM& LPIPS& Time & Params \\ \hline\hline
					RetinexNet~\cite{Chen2018Retinex} & 16.77 & 0.56 &0.47 & 0.06 & 0.44M \\
					RetinexNet~\cite{Chen2018Retinex}~+~BM3D & 17.91 & 0.73 &0.22 & 2.75 & 0.44M \\
					MBLLEN~\cite{lvmbllen}& 18.56 & 0.75 & 0.19 & \bf{0.05} & 0.31M  \\
					Ours-lightweight-1& 19.08 & 0.74  &0.17& \bf{0.05} & \bf{0.21M}  \\
					Ours-lightweight-2& 18.79 & 0.77  &0.21& \bf{0.05} & 0.25M  \\
					Ours-1& \bf{20.24} & 0.79  &\bf{0.14}& 0.06 & 0.88M  \\
					Ours-2& 19.48 & \bf{0.81}  &0.16& 0.06 & 0.92M  \\
					\hline\hline
					SID~\cite{seedark2018cvpr}& \bf{28.88} & \bf{0.79} & \bf{0.36} & 0.51 & 7.76M  \\
					Ours& 27.96 & 0.77 & \bf{0.36} & \bf{0.48} & \bf{0.88M}  \\
					\hline
	\end{tabular}}}} \\
\end{table}

\begin{figure*}[htbp]
	\begin{center}
		\begin{overpic}[width=1\textwidth]{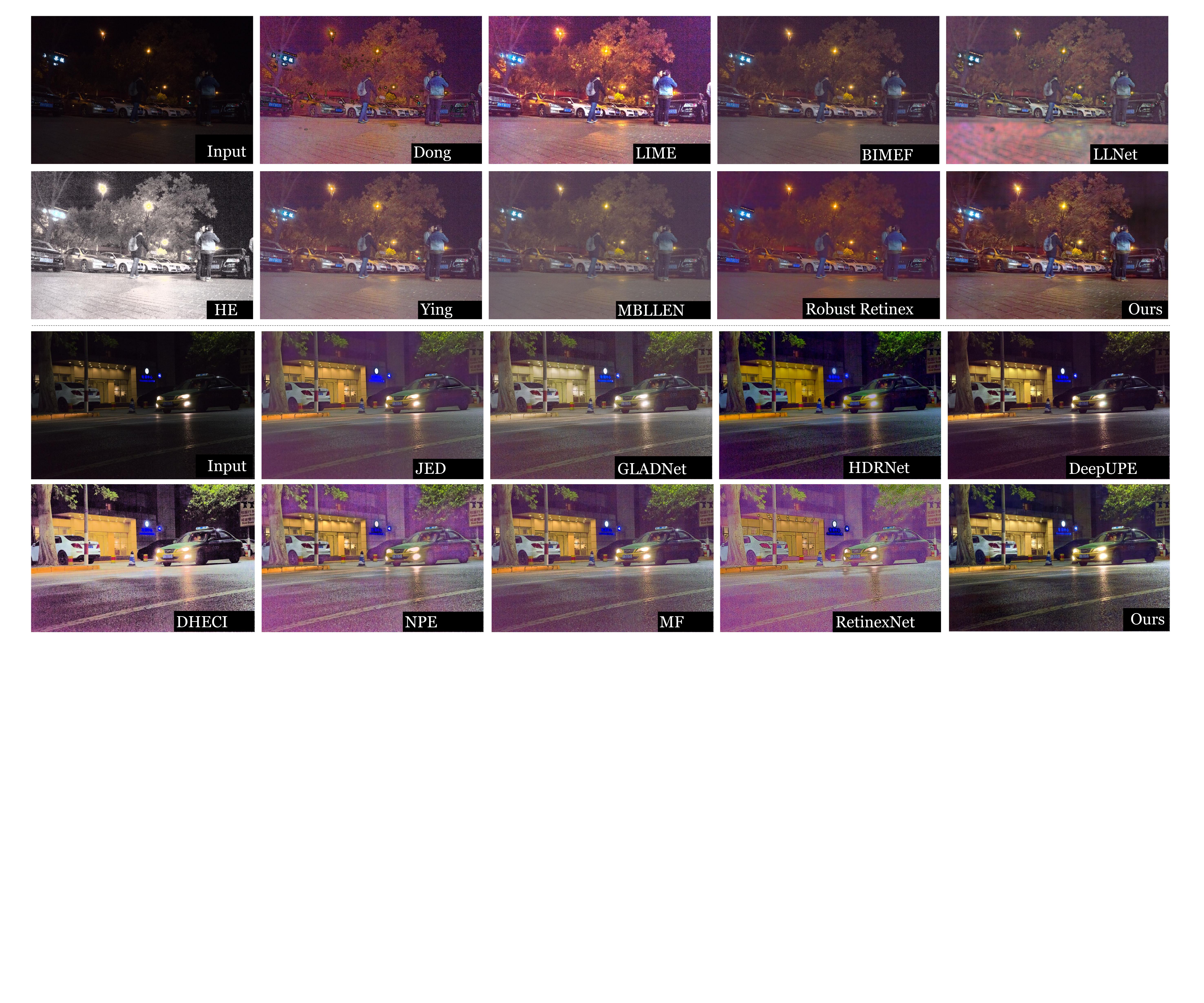}
			\put(182,206){\bf \color{white}\scriptsize \cite{dong2011fast}} 
			\put(284,206){\bf \color{white}\scriptsize \cite{guo2017lime}} 
			\put(381,206){\bf \color{white}\scriptsize \cite{ying2017bio}} 
			\put(480,206){\bf \color{white}\scriptsize \cite{lore2017llnet}} 
			
			\put(183.5,138){\bf \color{white}\scriptsize \cite{ying2017newiccv}} 
			\put(284,138){\bf \color{white}\scriptsize \cite{lvmbllen}} 
			\put(381,138){\bf \color{white}\scriptsize \cite{li2018structure}} 
			
			\put(182,69.8){\bf \color{white}\scriptsize \cite{ren2018joint}} 
			\put(284,69.8){\bf \color{white}\scriptsize \cite{wang2018gladnet}} 
			\put(381,69.8){\bf \color{white}\scriptsize \cite{gharbi2017deep}} 
			\put(480,69.8){\bf \color{white}\scriptsize \cite{wang2019underexposed}} 
			
			\put(86,3.5){\bf \color{white}\scriptsize \cite{nakai2013dheci}} 
			\put(180,3.5){\bf \color{white}\scriptsize \cite{wang2013naturalness}} 
			\put(284,3.5){\bf \color{white}\scriptsize \cite{fu2016mf}} 
			\put(383,3.5){\bf \color{white}\scriptsize \cite{Chen2018Retinex}} 
		\end{overpic}
	\end{center}
	\caption{Visual comparison of real low-light images, which are taken at night. Please zoom in for a better view.}
	\label{fig_nature_compare}
\end{figure*}
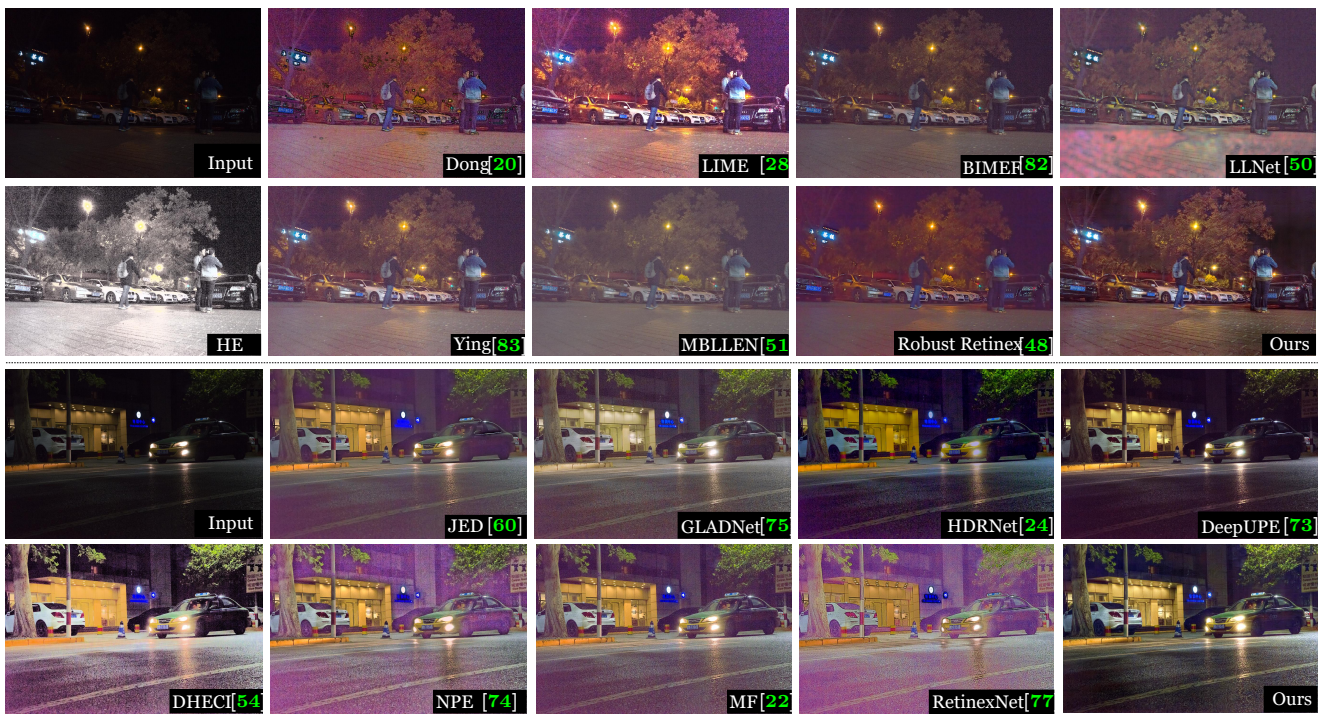

{\bf Natural low-light images.}
We first compare our method with state-of-the-art methods on natural low-light images and the representative visual comparison results are shown in Figure~\ref{fig_nature_compare}. Our method surpasses other methods in two key aspects. On the one hand, our method can restore vivid and natural color to make the enhancement results more realistic. In contrast, Retinex-based methods (such as RetinexNet~\cite{Chen2018Retinex} and LIME~\cite{guo2017lime}) will cause different degrees of color distortion. On the other hand, our method is able to recover better contrast and more details. This improvement is especially evident when compared with LLNet~\cite{lore2017llnet}, BIMEF~\cite{ying2017bio} and MBLLEN~\cite{lvmbllen}.

{\bf User study.}
We invite $100$ participants to attend a user study to test the subjective preference of low-light image enhancement methods. We randomly select $20$ natural low-light image cases and enhance them using five representative methods. For each case, the input data and the five enhanced results will be shown to the participants at the same time. We then ask the participants to rank the quality of the five enhancements from 1 (best) to 5 (worst) in terms of recovery of brightness, contrast, and color. We also provide zoom-in function to let participants to check details like texture and noises controls. The other four methods used besides ours in this study are DHECI~\cite{nakai2013dheci}, DeepUPE~\cite{wang2019underexposed}, LIME~\cite{guo2017lime} and Robust~\cite{li2018structure}.

%

\begin{figure}[tbp]
	\begin{center}
		\begin{overpic}[width=0.48\textwidth]{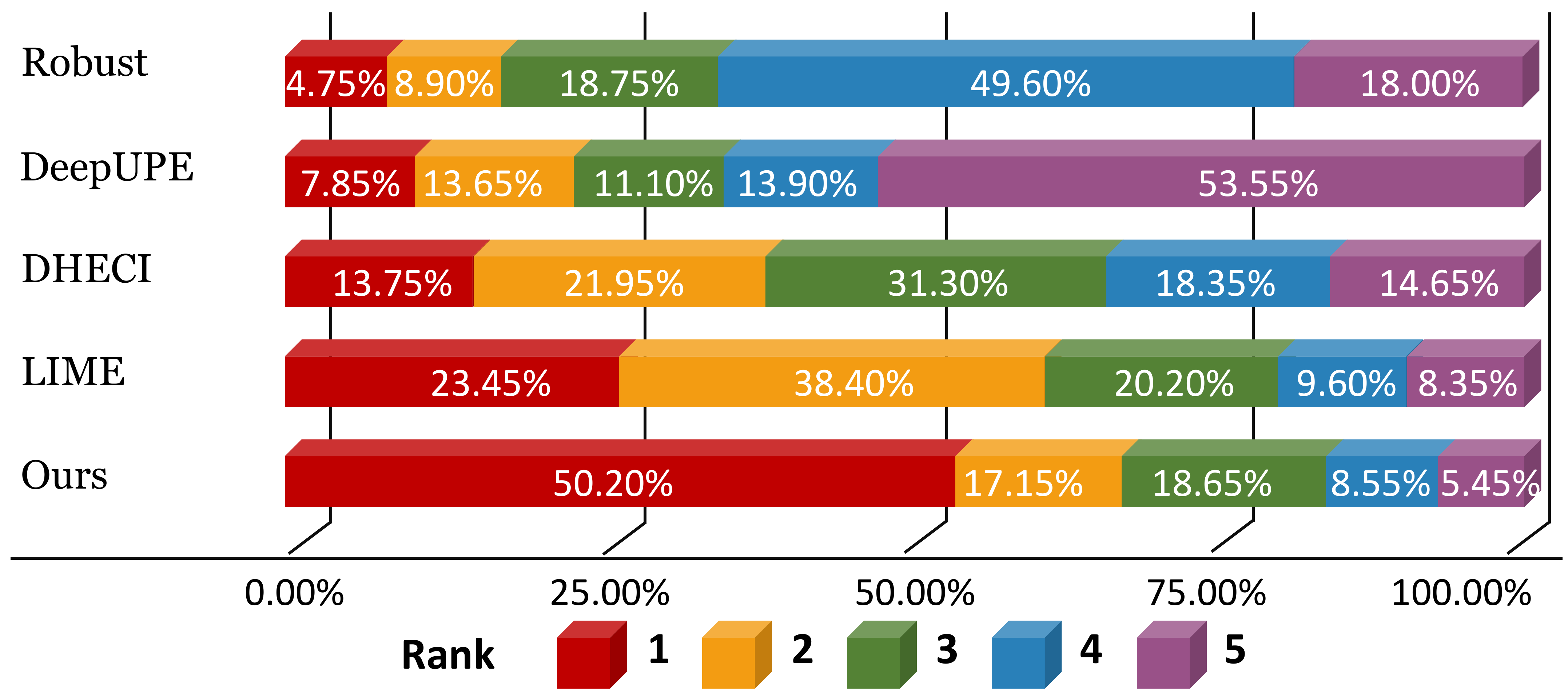}
			\put(26,93.8){\bf \color{black}\scriptsize \cite{li2018structure}} 
			\put(29.5,78){\bf \color{black}\scriptsize \cite{wang2019underexposed}} 
			\put(26,63){\bf \color{black}\scriptsize \cite{nakai2013dheci}} 
			\put(23,47.1){\bf \color{black}\scriptsize \cite{guo2017lime}} 
		\end{overpic}
	\end{center}
	\caption{Rating distribution of the user study.}
	\label{fig_user_study}
\end{figure}

Figure~\ref{fig_user_study} shows the rating distribution of the user study. Our method receives more ``best" ratings, which shows that our results are more preferred by human subjects.

\begin{figure*}[t]
	\begin{center}
		\includegraphics[width=1\textwidth]{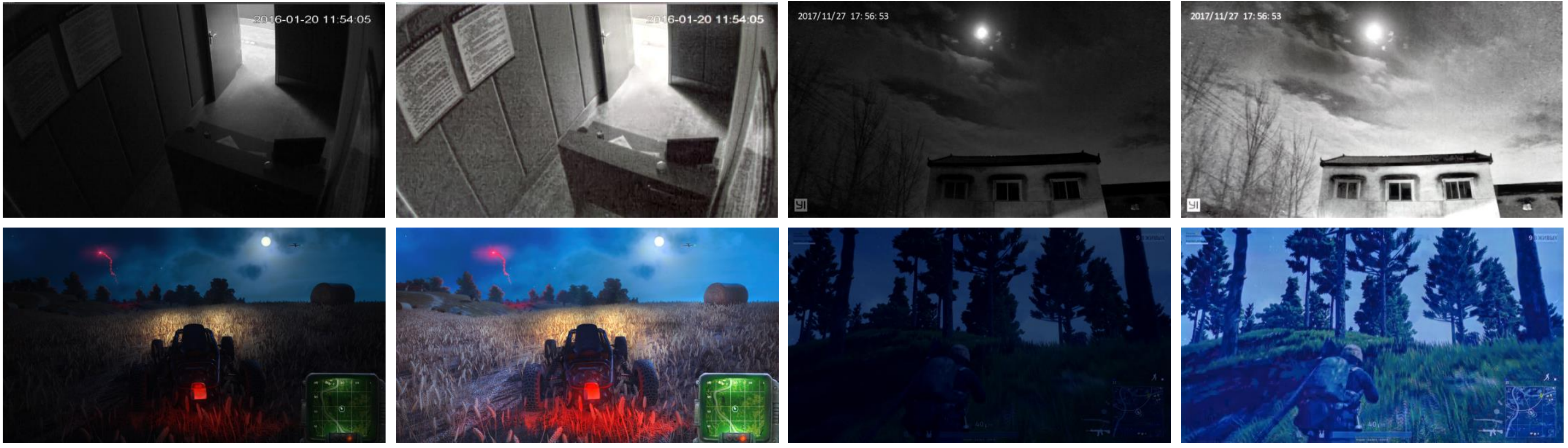}
	\end{center}
	\vspace{-0.1cm}
	\caption{Generalizing our method to enhance (upper) monochrome surveillance scenes and (bottom) nighttime game scenes.}
	\label{fig_game_compare}
\end{figure*}

\begin{figure*}[t]
	\begin{center}
		\includegraphics[width=1\textwidth]{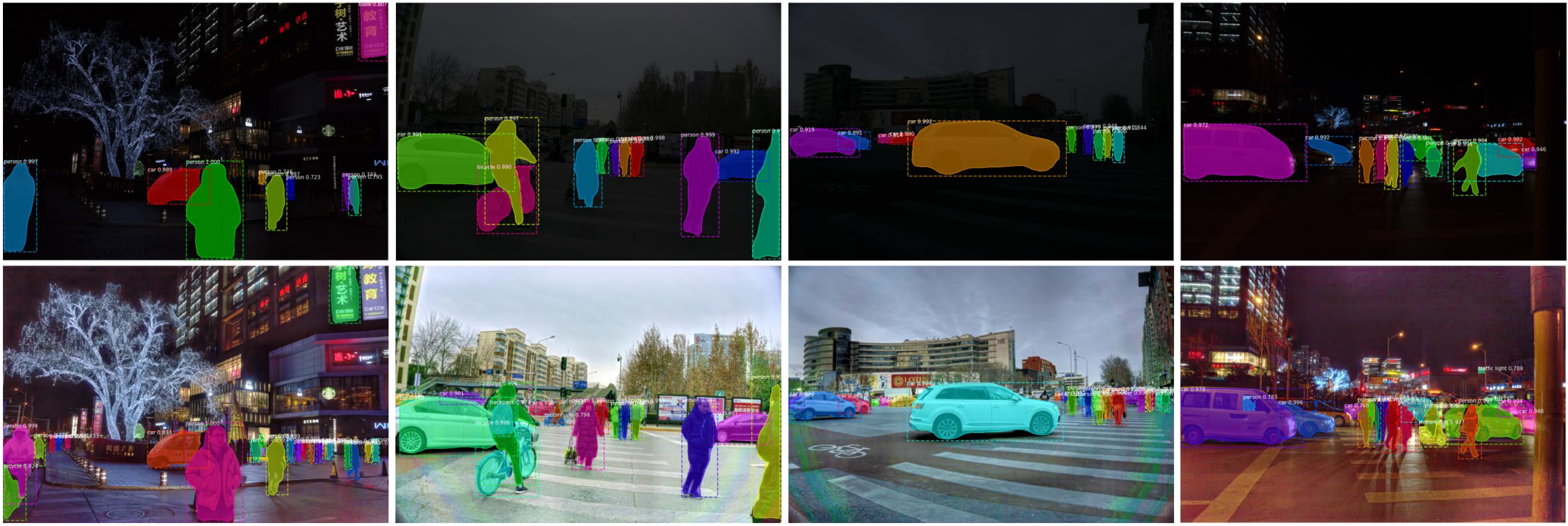}
	\end{center}
	\vspace{-0.1cm}
	\caption{After processing the low-light scene (upper row) with our method, the performance of both object detection and instance segmentation are greatly improved (bottom row).}
	\label{fig_detection}
\end{figure*}

\begin{figure}[htbp]
	\begin{center}
		\begin{overpic}[width=0.48\textwidth]{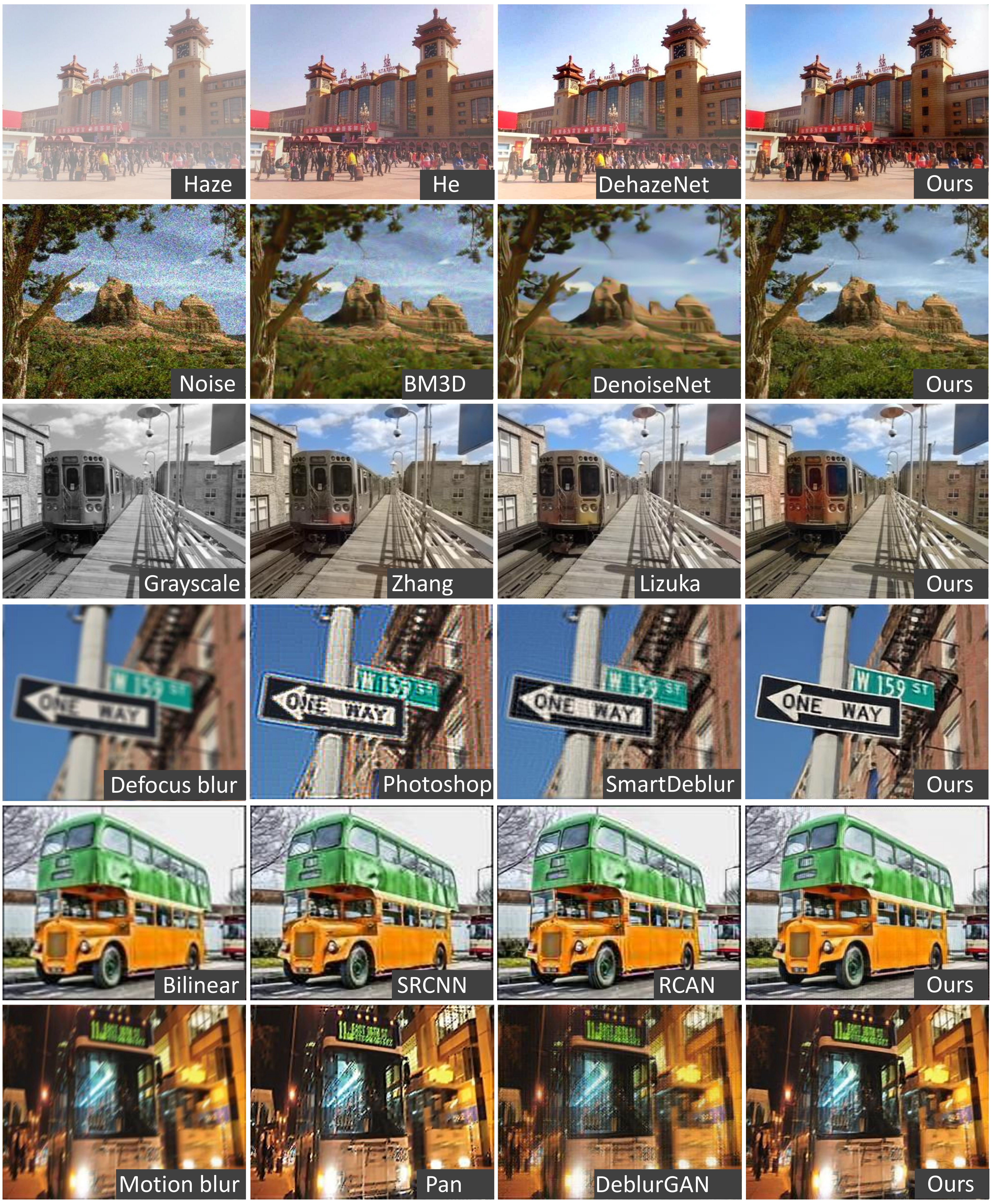}
			\put(110.5,242.5){\bf \color{white}\tiny \cite{he2011single}} 
			\put(170,242.5){\bf \color{white}\tiny \cite{cai2016dehazenet}} 
			
			\put(111,194.5){\bf \color{white}\tiny \cite{dabov2006image}} 
			\put(170.5,194.5){\bf \color{white}\tiny \cite{remez2017deep}} 
			
			\put(108.5,147){\bf \color{white}\tiny \cite{zhang2016colorful}} 
			\put(168.5,147){\bf \color{white}\tiny \cite{iizuka2016let}} 
			
			\put(111,51){\bf \color{white}\tiny \cite{dong2014learning}} 
			\put(170.5,51){\bf \color{white}\tiny \cite{zhang2018rcan}} 
			
			\put(110,3.3){\bf \color{white}\tiny \cite{pan2016blind}} 
			\put(170,3.3){\bf \color{white}\tiny \cite{DeblurGAN}} 
		\end{overpic}
	\end{center}
	\caption{Visual comparison of several low-level vision tasks. From top to bottom: dehazing, denoising, colorization, defocus deblurring, super-resolution ($\times 2$) and motion deblurring. }
	\label{fig_lowlevel}
\end{figure}
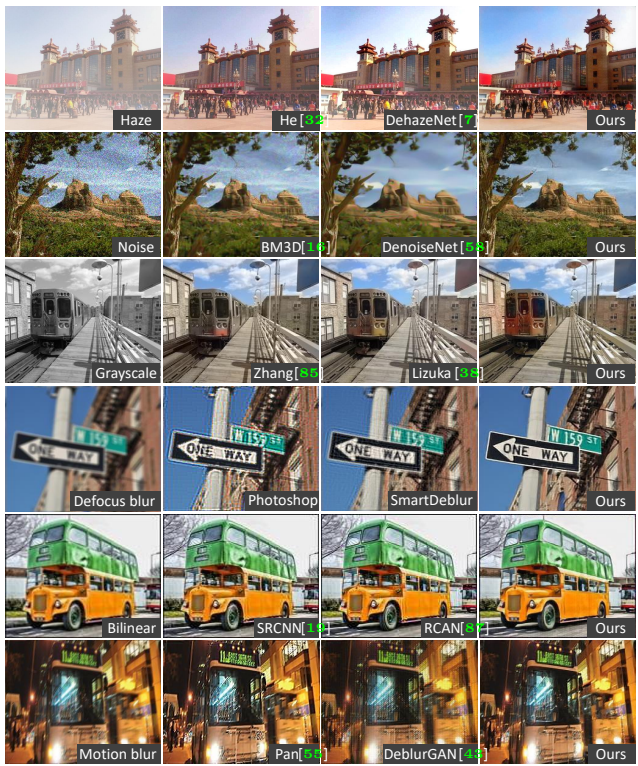

{\bf Generalization study.} To prove the robustness of our method, we directly apply our trained model to enhance some specific types of low-light scenes (such as monochrome surveillance and game night scenes) that are unseen in the training dataset. Figure~\ref{fig_game_compare} shows the enhancement results. The results demonstrate that our method is robust and effective for general low-light image enhancement tasks. Besides, we also show that our approach is beneficial to some high-level tasks in low-light scenes, such as object detection and instance segmentation, as shown in Figure~\ref{fig_detection}. The performance of Mask-RCNN~\cite{matterport_maskrcnn_2017,He2017MaskR} has been improved a lot by using our method in a pre-processing stage without any fine-tuning.
Besides, we have tested end-to-end training our multi-branch network for many other low-level computer vision tasks and demonstrated the effectiveness. Visual examples on denoising, dehazing, deblurring, etc., are shown in Figure~\ref{fig_lowlevel}.

\subsection{Ablation Study}
In this section, we quantitatively evaluate the effectiveness of different components in our method based on our synthetic low-light dataset. Table~\ref{tab_control} reports the accuracy of the presented change in terms of PSNR and SSIM~\cite{wang2004image}. Note that the Reinforce-Net is not considered in this study.

{\bf Loss functions.} We mainly evaluate the loss function of the Enhancement-Net, as shown in Table~\ref{tab_control} (row 2-5). We use \textit{mse} as the naive loss function under condition 2. The results show that the quality of enhancement is improving by containing more loss components.

\begin{table}[t]
	\caption{Ablation study. This table reports the performance under each condition based on the synthetic low-light dataset. In this table, "w/o" means without.} 
	\label{tab_control}
	\centering{\scalebox{0.95}{\scalebox{1}{\begin{tabular}{l|cccc}
					\hline
					Condition  & PSNR& SSIM \\ \hline\hline
					1. default configuration &  {\bf 20.84} & {\bf 0.82} \\
					\hline\hline
					2. w/o $\mathcal{L}_{eb}$, w/o $\mathcal{L}_{es}$, w/o $\mathcal{L}_{ep}$, w/o $\mathcal{L}_{er}$ &  19.36 & 0.73 \\
					3. with $\mathcal{L}_{eb}$, w/o $\mathcal{L}_{es}$, w/o $\mathcal{L}_{ep}$, w/o $\mathcal{L}_{er}$ &  20.01 & 0.76 \\
					4. with $\mathcal{L}_{eb}$, with $\mathcal{L}_{es}$, w/o $\mathcal{L}_{ep}$, w/o $\mathcal{L}_{er}$ &  19.92 & 0.78 \\
					5. with $\mathcal{L}_{eb}$, with $\mathcal{L}_{es}$, with $\mathcal{L}_{ep}$, w/o $\mathcal{L}_{er}$ &  20.58 & 0.81 \\
					\hline\hline
					6. w/o Attention-Net, w/o Noise-Net &  19.12 & 0.71 \\
					7. with Attention-Net, w/o Noise-Net &  20.66 & 0.80 \\
					\hline\hline
					8. branch number $\times 1$ ($5$) &  20.66 & 0.79 \\
					9. branch number $\times 3$ ($15$) &  20.83 & {\bf 0.82} \\
					\hline
	\end{tabular}}}} \\
\end{table}

{\bf Network structures.} As shown in Table~\ref{tab_control} (row 6-7), we evaluate the effectiveness of different network components. Similar to the loss function, the results demonstrate that more components of our network will result in better performance.


{\bf Number of branches.} We analyze the effect of different branch numbers (model size) on the network performance, as shown in Table~\ref{tab_control} (row 8-9). Obviously, the increase of model size will not always improve performance, so we set $10$ branches as the default configuration.

%

\subsection{Unsatisfying Cases}
\begin{figure}[t]
	\begin{center}
		\includegraphics[width=0.48\textwidth]{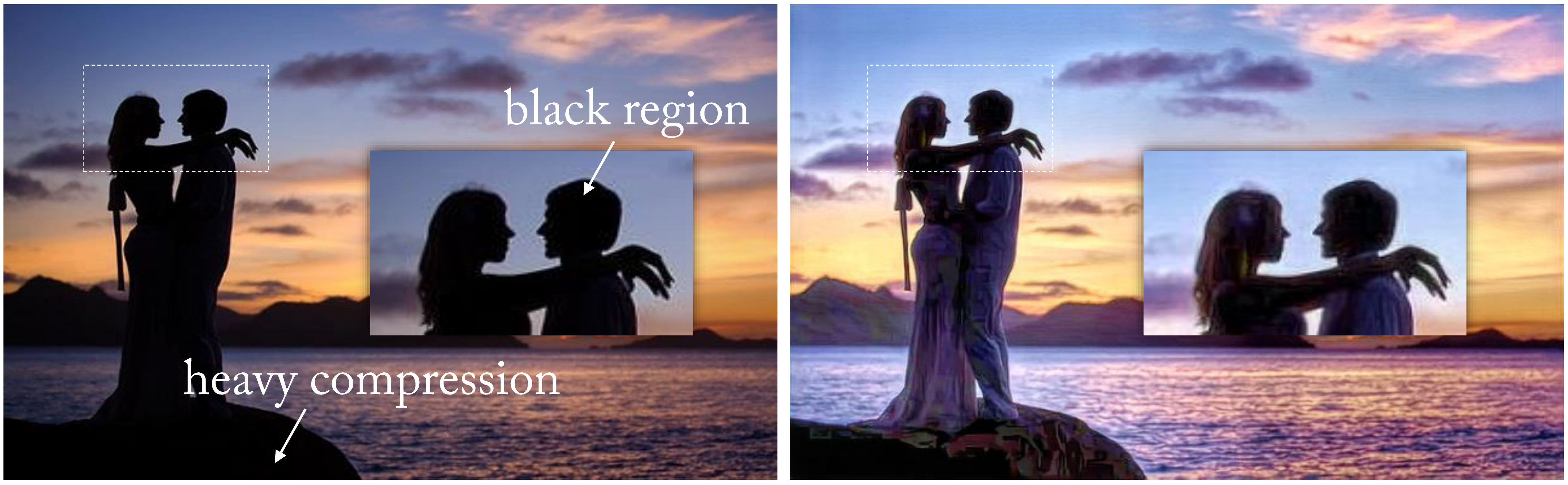}
	\end{center}
	\caption{This image has totally dark region where textures are lost and heavy compression. These cause issues in the enhancement result.}
	\label{fig_failurecase}
\end{figure}

Figure~\ref{fig_failurecase} presents a case where our method performs not perfectively. Our method fails to recover the face details on the top image, as some parts of the face are totally dark. Another issue is the blocking artifacts due to heavy image compression.

%

\section{Conclusion}
This paper proposes an attention-guided enhancement solution for low-light image enhancement. We design a multi-branch network to handle enhance the brightness and handle the noise simultaneously. The key is to use the proposed ue-attention map and noise map to guide the enhancement in a region adaptive manner. We also propose a low-light image simulation pipeline and build a large-scale low-light enhancement benchmark dataset for model training and evaluation.
Extensive experiments demonstrate that our solution outperforms state-of-the-art methods by a large margin.
As for future direction, extending the proposed method to low-light video enhancement is of our interest.

%
%


\bibliographystyle{spmpsci}      
\bibliography{egbib.bib}   

\end{document}